\documentclass[a4paper,12pt]{article}
\pdfoutput=1
\usepackage{graphicx, rotating}
\usepackage{hyperref}
\usepackage{slashed}
\usepackage{ifpdf}

\ifx\pdfoutput\undefined
\usepackage[dvips,bookmarks=false]{hyperref}	
\else
\usepackage{hyperref}	
\fi
\hypersetup{colorlinks,bookmarksopen,bookmarksnumbered,citecolor=verdes,
linkcolor=blus,pdfstartview=FitH,urlcolor=rossos}
\def\hhref#1{\href{http://arxiv.org/abs/#1}{#1}} 

\usepackage{multicol,amsmath}
\usepackage{color}
\definecolor{rosso}{cmyk}{0,1,1,0.4}
\definecolor{rossos}{cmyk}{0,1,1,0.55}
\definecolor{rossoc}{cmyk}{0,1,1,0.2}
\definecolor{blu}{cmyk}{1,1,0,0.3}
\definecolor{blus}{cmyk}{1,1,0,0.6}
\definecolor{bluc}{cmyk}{1,1,0,0.1}
\definecolor{verde}{cmyk}{0.92,0,0.59,0.25}
\definecolor{verdec}{cmyk}{0.92,0,0.59,0.15}
\definecolor{verdes}{cmyk}{0.92,0,0.59,0.4}

\font\tenrsfs=rsfs10 at 12pt
\font\sevenrsfs=rsfs7
\font\fiversfs=rsfs5
\newfam\rsfsfam
\textfont\rsfsfam=\tenrsfs
\scriptfont\rsfsfam=\sevenrsfs
\scriptscriptfont\rsfsfam=\fiversfs
\def\mathscr#1{{\fam\rsfsfam\relax#1}}

\newcommand{\fig}[1]{~\ref{fig:#1}}
\newcommand\DM{{\rm  DM}}
\newcommand{\eq}[1]{~{\rm (\ref{eq:#1})}}

\newcommand{\GeV}{\,{\rm GeV}}
\newcommand{\TeV}{\,{\rm TeV}}

\def\circa#1{\,\raise.3ex\hbox{$#1$\kern-.75em\lower1ex\hbox{$\sim$}}\,}

\newcommand{\pl}{p\hspace{-4.2pt}{\scriptstyle/}}

\newcommand{\beq}{\begin{equation}}
\newcommand{\eeq}{\end{equation}}

\def\eq#1{~(\ref{eq:#1})}
\def\circa#1{\,\raise.3ex\hbox{$#1$\kern-.75em\lower1ex\hbox{$\sim$}}\,}
\makeatletter

\def\art{\@ifnextchar[{\eart}{\oart}}
\def\eart[#1]#2#3#4#5#6{{\rm #2}, {#3 #4} {\rm (#6) #5} [arXiv:\-{\hhref{#1}}]}
\def\hepart[#1]#2{{\rm #2, arXiv:\-\hhref{#1}}}
\newcommand{\oart}[5]{{\rm #1}, {#2 #3} {\rm (#5) #4}}

\newcounter{alphaequation}[equation]
\def\thealphaequation{\theequation\hbox to
0.6em{\hfil\alph{alphaequation}\hfil}}
\def\eqnsystem#1{
\def\@eqnnum{{\rm (\thealphaequation)}}
\def\@@eqncr{\let\@tempa\relax \ifcase\@eqcnt \def\@tempa{& & &} \or
\def\@tempa{& &}\or \def\@tempa{&}\fi\@tempa
\if@eqnsw\@eqnnum\refstepcounter{alphaequation}\fi
\global\@eqnswtrue\global\@eqcnt=0\cr}
\refstepcounter{equation} \let\@currentlabel\theequation \def\@tempb{#1}
\ifx\@tempb\empty\else\label{#1}\fi
\refstepcounter{alphaequation}
\let\@currentlabel\thealphaequation
\global\@eqnswtrue\global\@eqcnt=0 \tabskip\@centering\let\\=\@eqncr
$$\halign to \displaywidth\bgroup \@eqnsel\hskip\@centering
$\displaystyle\tabskip\z@{##}$&\global\@eqcnt\@ne
\hskip2\arraycolsep\hfil${##}$\hfil& \global\@eqcnt\tw@\hskip2\arraycolsep
$\displaystyle\tabskip\z@{##}$\hfil
\tabskip\@centering&\llap{##}\tabskip\z@\cr}
\def\endeqnsystem{\@@eqncr\egroup$$\global\@ignoretrue} \makeatother

\newcommand{\SU}{\rm SU}

\begin{document}

\begin{center}
IFUP-TH/2010-24\hfill
CERN-PH-TH/2010-179

\bigskip\bigskip\bigskip

{\huge\bf\color{magenta}
 Weak Corrections are Relevant for\\[3mm]
Dark Matter Indirect Detection}\\

\medskip
\bigskip\color{black}\vspace{0.6cm}
{\bf
\large Paolo Ciafaloni$^{(a)}$, Denis Comelli$^{(b)}$, Antonio Riotto$^{(c,d)}$,\\
Filippo Sala$^{(e,f)}$, Alessandro Strumia$^{(c,e,g)}$, Alfredo Urbano$^{(h)}$}
\\[7mm]
{\it $^a$   INFN - Sezione di
Lecce, Via per Arnesano, I-73100 Lecce, Italy} \\
{\it $^b$  INFN - Sezione di Ferrara, Via Saragat 3, I-44100 Ferrara, Italy}\\
{\it $^c$ CERN, PH-TH, CH-1211, Geneva 23, Switzerland}\\
{\it $^d$ INFN, Sezione di Padova, Via Marzolo 8, I-35131, Padova, Italy}\\
{\it $^e$ Dipartimento di Fisica dell'Universit{\`a} di Pisa and INFN, Italy}\\
{\it $^f$ Scuola Normale Superiore, Piazza dei Cavalieri 7, I-56126 Pisa, Italy}\\
{\it $^g$  National Institute of Chemical Physics and Biophysics, Ravala 10, Tallin, Estonia}\\
{\it $^h$   Dipartimento di Fisica, Universit\`a di Lecce and INFN - Sezione di
Lecce, \\Via per Arnesano, I-73100 Lecce, Italy} \\

\bigskip\bigskip\bigskip\bigskip

{
\centerline{\large\bf Abstract}
\begin{quote}
The computation of the energy spectra of Standard Model  particles originated from the annihilation/decay of dark matter  particles is of primary importance in indirect searches of dark matter.
We compute how the inclusion of 
electroweak corrections significantly alter such spectra when the  mass $M$ of  dark matter particles  is larger than the electroweak scale:
soft electroweak  gauge bosons are copiously radiated opening new channels in the final
states which otherwise would be  forbidden if such corrections are neglected. All stable particles
are therefore  present in the final spectrum, independently of the primary channel of
dark matter annihilation/decay.
Such corrections are model-independent.

\end{quote}}

\end{center}

\newpage 
\tableofcontents

\section{Introduction}
There are overwhelming cosmological and astrophysical evidences that our universe contains
a sizable amount of Dark Matter (DM), {\it i.e.} a component which clusters at small scales.
While its abundance is known rather well in terms of the critical energy density, $\Omega_{\rm DM}h^2=0.110\pm 0.005$~\cite{wmap}, its nature is still a mistery. Various considerations point
towards the possibility that DM is made of neutral particles.
If DM is composed by particles whose mass and interactions  are dictated by  physics in the electroweak
energy range, its abundance is likely to be fixed by the thermal freeze-out phenomenon within the
standard Big-Bang theory. DM particles, if present in thermal abundances in the early
universe, annihilate with one another so that a predictable number of
them remain today.  The relic density of these particles comes out to
be:
\begin{equation}\label{eq:Omega}
\frac{\Omega_{\rm DM} h^2}{0.110} \approx\frac{3 \times 10^{-26} {\rm cm}^3/{\rm sec}}{ \langle \sigma v \rangle_{\rm ann}},
\end{equation}
where  $\langle \sigma v \rangle_{\rm ann} $ is the (thermally-averaged) cross annihilation cross
sections.
A  weak interaction strength provides  the abundance in the right range. This numerical coincidence
represents the main  reason why it is generically believed that DM is
made of weakly-interacting particles with a  mass in the range
$(10^2-10^4)$ GeV. There are several ways to search for such DM
candidates.
 If they are light enough, they might reveal themselves in {\em particle colliders}, such as
the LHC, as missing energy in an event.  In that case one
knows that the particles live long enough to escape the detector, but
it will still be unclear whether they
are long-lived enough to be the DM \cite{dmlhc}.  Thus
complementary  experiments are needed. In {\em direct
detection} experiments, the DM particles  elastically scatter off of a nucleus in the
detector, and a number of experimental signatures of the interaction
can be detected \cite{dmdirect}. In {\em indirect searches}  DM
annihilations or decays
around the Milky Way can produce Standard Model (SM)  particles that
decay into
 $e^{\pm}, p,\overline{p},\gamma$ and $\overline{d}$ , producing an excess in their cosmic ray fluxes.
Present observations are approaching the sensitivity needed to probe
 the annihilation cross section suggested by cosmology, eq.\eq{Omega}.

 Furthermore, this topic recently attracted interest because the PAMELA experiment \cite{pamela}  observed an unexpected
rise with energy of the $e^+/(e^+ +e^-)$ fraction in cosmic rays, suggesting the existence of
a new positron component. The sharp rise might  suggest that the new component may be visible also
in the $(e^+ + e^-)$ spectrum: although the peak hinted by previous ATIC data \cite{atic} is not confirmed,
the FERMI \cite{fermi} and HESS \cite{hess} observations still demonstrate a deviation from the naive power-law
spectrum, indicating an excess compared to conventional background predictions of cosmic ray
fluxes at the Earth.
While the current excesses might be either due  to a new astrophysical component, such as a nearby
pulsar \cite{pul}, or to some experimental problem, it could be produced by DM with a cross section a few orders of magnitude larger than in eq.\eq{Omega}, maybe thanks to a Sommerfeld enhancement~\cite{PAMth,Arcano}.

\smallskip

In any case, it is undeniable  that nowadays indirect search
of DM is a fundamental topic in astroparticle physics, both from the theoretical and experimental point of view. Computing
 the energy
spectra of the stable SM particles that are
present in cosmic
rays and might originate from DM annihilation/decay is therefore of primary
importance.

\medskip

The key point of this paper
is to show that
electroweak radiative corrections have a sizable impact on  the
energy spectra of SM particles originated from the annihilation/decay of DM particles with
mass  $M$ somehow larger than the electroweak scale. The reason is in fact simple
and should be familiar to readers working in collider physics: at
energies much higher than
 the weak scale (in our case the mass $M$ of the DM)
soft electroweak  gauge bosons are copiously radiated from highly energetic objects
(in our case the initial products of the DM annihilation/decay). This emission is enhanced by
$\ln M^2/M_W^2$ when collinear divergences are present and $\ln^2 M^2/M_W^2$
when both collinear and infrared divergences are present \cite{cc}.
These logarithmically enhanced terms can be computed in a model-independent way through the well known partonic techniques
based on the Collinear Approximation (CA).
Our work will involve generalizing the partonic splitting functions to massive partons, because
our `partons' include the $W,Z$ bosons.

Putting these technical details aside, what is important is that the  emission of gauge bosons
changes significantly some final energy spectra. Indeed, suppose that DM
annihilates into a pair of leptons. The emitted  gauge bosons give hadrons
(resulting in a $\bar p$ flux) and mesons (giving a significant extra amount of photons via $\pi^0\to\gamma\gamma$).
The total energy gets distributed  among a large number
of lower energy particles, thus  enhancing the signal in the lower energy region (say,  $(10-100)$ GeV),
that is measured by present-day experiments,  like PAMELA.

\medskip

This paper will be inevitably rather technical and therefore we have
decided to defer as many as
possible technicalities to the various Appendices. To diminish the burden, we present
qualitative considerations in Section 2 and outline the quantitative computation in Section 3.
The reader interested in the final results may jump directly to Section
4 where our findings are presented. Conclusions are presented in Section 5.
In Appendix A we discuss EW evolution equations, in Appendix B we derive
parton splitting functions for massive partons, and in Appendix C we list all splittings among SM particles,
including the effect of the top Yukawa coupling.


\section{Qualitative discussion}
As mentioned in the Introduction,  the presence of DM is probed indirectly by detecting the energy
spectra of stable particles ($p^\pm,e^\pm,\nu,\gamma,\bar{d}$). At a first
sight, since electroweak
 radiative corrections are expected to
be small ---  weak interactions are weak, after all ---
they might seem to play no role in the DM indirect searches. At the typical weak
scale of ${\cal O}(100)$ GeV, radiative corrections produce relative effects
of ${\cal O}(0.1)$\%.  For instance, this was  the case for experiments that took place at
the LEP collider.
However at energies of the order of the TeV scale, like those probed
at the LHC,  things are
 different: electroweak  radiative corrections
can reach the ${\cal O}(30)$\% level  \cite{LHC} and they grow with
energy, eventually calling for a  resummation of higher order effects \cite{NLL}.
In a nutshell, what happens at energies much higher than the weak scale is that
soft electroweak  gauge bosons are copiously radiated from highly energetic objects
that undergo a scattering with high invariant mass.
 This is much the
same as photon, or gluon, radiation whenever the hard scale is such that
the $W,Z$ masses can be safely taken to be very small.
Important differences with respect to unbroken gauge theories
like QCD and QCD arise in the case of a (spontaneously) broken theory like
the EW sector of the SM.
It was found \cite{BN}  that  in hard processes with at least two relativistic
non abelian charges,  effective
infrared divergencies
that are manifest as double log corrections
($\alpha_2\ln^2M^2/M_W^2$) appear.
They are not present in QED and QCD and  this effect has been baptized
``Bloch-Nordsieck Theorem Violation'' \cite{BN}.
We refer the reader to the relevant literature
\cite{BN,long,generic}
for details. In the case at hand, since the initial DM particles are
nonrelativistic, radiation related to the initial legs
does not produce log-enhanced terms. Therefore, we only need to examine
soft EW radiation related to the final state particles.

The hard scale in the case we examine here
is provided by the DM mass $M\circa{>}1$ TeV while the soft scale is
the typical energy where the spectra of the final products of DM decay/annihilation
 are measured, $E\circa{<} 100$ GeV.
\begin{figure}
\begin{center}
$$\includegraphics[width=0.5\textwidth]{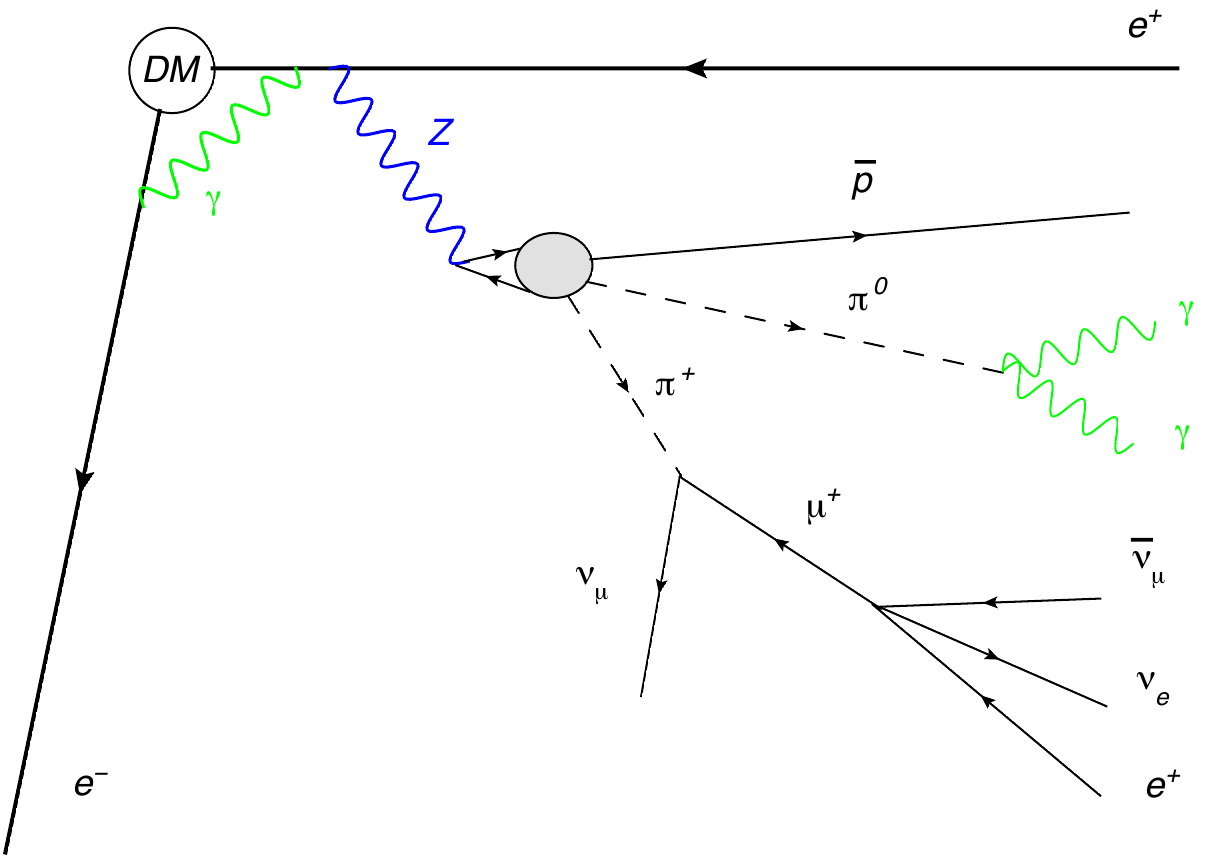}$$
\caption{\em DM annihilation/decay initially
  produces a hard positron-electron pair. The spectrum of the hard objects
is altered by electroweak virtual corrections (green photon line) and real $Z$
  emission. The $Z$  decays hadronically through a $q\bar{q}$ pair and
  produces a great number
of much softer objects, among which an antiproton and two
  pions; the latter
 cascade decay to softer $\gamma$s and leptons.
  \label{fig:cascade}}
\end{center}
\end{figure}
Even bearing in mind that weak interactions are not so weak at the TeV scale,
one might wonder whether such ``strong'' electroweak  effects are relevant for
measurements with
 uncertainties
 very far from the precision reachable by ground-based
experiments at colliders. In this context, and in view of our ignorance
about the physics responsible for DM cross sections, it might seem that
even a ${\cal O}(30)$\% relative effect should
have a minor impact.
 This is by no means the case:  including electroweak  corrections has
a huge impact on the measured energy spectra from DM decay/annihilation.
There are two basic reasons for this   rather surprising result.
\begin{itemize}
\item In the first place,  since
energy is conserved, but the total number of particles is not, because of
 electroweak   radiation a small number of highly energetic particles is
converted into a great number of low energy particles, thus enhancing the
low energy ($\circa{<} 100$ GeV) part of the spectrum, which is the one of
relevance from the experimental point of view.
\item Secondly, and perhaps more importantly:  since all SM particles are charged under
the $\SU(2)_L\otimes {\rm U}(1)_Y$ group,  including electroweak corrections opens new channels in the final
states which otherwise would be  forbidden if such corrections are neglected. In other
words, since electroweak corrections link all SM particles, all stable particles
will be present in the final spectrum, independently of the primary
annihilation  channel considered.
\end{itemize}
To illustrate these facts, consider for instance a heavy DM annihilation producing an
electron-positron pair, see  Fig. \ref{fig:cascade}.
Clearly, as long as one does not  take into account weak
interactions, only the leptonic channel is active and no antiproton is
present in the final products. However,  at very high
energies there is a probability of order unity that the positron radiates a $Z$ or a $W$.
While the spectrum of the
hard positron is not much altered by virtual and real
radiative corrections (see
\cite{CF}), the $Z$ radiation opens the hadronic channel: for instance,
antiprotons
are produced in the $Z$ decay. Moreover, also a large number of pions are
produced, which in turn decay to photons ($\pi^0\to\gamma\gamma$) and to
low energy positrons (through the chain $\pi^+\to\mu^++X\to e^++X$).
At every step, energy is degraded.
Because of the large multiplicity in the final states, the total $Z$ energy
(already smaller than the hard $M$ scale) is distributed among a large number
of objects, thus greatly enhancing the signal in the $(10-100)$ GeV region
that is measured by present-day experiments,  like PAMELA.

The various processes of radiation are
described by fragmentation
functions $D^{\rm EW}(x,\mu^2)$ that evolve with the 
 energy scale $\mu^2$
according to a set of integro-differential electroweak  equations  \cite{EvEq}.
 When a value of virtuality of the order of the weak scale $\mu=M_W$ is reached, the $Z$ boson
is on shell and decays. The subsequent QCD showering may be described with QCD traditional  MonteCarlo (MC) generator tools, like  PYTHIA.

At tree level, the spectra of hard objects emerging from DM annihilation  are  simply proportional to $\delta(1-x)$,
where $x$
is the fraction of center-of-mass  energy carried
by a given particle. Once electroweak
corrections are switched on and ${\cal O}(\alpha_2)$
  virtual and real corrections
are calculated, the spectra $D^{\rm EW}(x,\mu^2=M_W^2)$ generically
contains terms enhanced by log terms of the form
$\alpha_2 \ln^2M^2/M_W^2$ and $\alpha_2 \ln M^2/M_W^2$.
The presence of logarithmically enhanced terms
 is well-known in the literature both in the case of electroweak interactions
\cite{BN,EvEq} than for
 strong interactions through the Dokshitzer-Gribov-Lipatov-Altarelli-Parisi (DGLAP) equations \cite{DGLAP} and is related to
regions of phase space where the internal propagators become singular.
In general the single log term is generated when two  partons become  collinear,
while the double log arises when they are  soft and collinear at the same time.
 The double log-enhanced contributions cancel
 in  QED and QCD  for  physical scattering processes (Block-Nordsieck theorem \cite{bnt}),
 while   they are    present in massive gauge theories \cite{BN}.

The physical picture that arises is therefore the following:
highly non relativistic DM particles annihilate, producing a
particle-antiparticle pair belonging to the SM spectrum. This pair has a
very high invariant mass, therefore it stars radiating photons and gluons,
but also weak gauge bosons.
 The presence of collinear and/or infrared
singularities allows to factorize leading logarithmic electroweak
corrections with a
probabilistic interpretation very similar to DGLAP equations, see
Sec. 3.  The exchange of virtual and emission of real
 electroweak bosons lead to the appearance in the final spectrum of all the stable
SM particles, not only the ones initially emitted by the DM
annihilation.
 Indeed, the higher is the mass of the DM, the more democratically distributed the final spectrum of DM particles is.
Therefore, including electroweak corrections alters significantly the final spectrum of particles
stemming from DM decay/annihilation and this has a large impact on indirect searches of DM.

\medskip

Let us close this Section by recalling that, while in this paper we only
consider DM annihilation/decay to two body final states, our approach is
more general and model independent. Indeed, the only assumptions we make
are that
the physics up to the DM mass scale $M$ is described by the SM and
that the SM may be eventually  extended by interactions that preserve
 $\SU(2)_L\otimes {\rm U}(1)_Y$ gauge invariance.
While these assumptions exclude cases like the ones considered in
Refs. \cite{Arcano}
and \cite{Pasquale} where gauge non invariant interactions where considered\footnote{The analysis of the Infrared virtual corrections to gauge non invariant amplitudes, i.e. amplitudes proportionals to the higgs vev, has been recently performed in
 \cite{anomalous}. 
}, a large number of models can be examined with the
techniques we describe here.
For instance, let us  extend the SM by adding a very heavy scalar $S$ that
interacts with the SM Higgs ($H$) and leptons ($L,E$),  through an
effective operator $SLEH$. Then, the dominant decay of the scalar is a
three body decay, since the two body decay $S\to LE$ is suppressed
by a relative factor $M_W^2/M^2$. The framework described here
applies as well, albeit with the additional complication that the three body
decay with respect to which one factorizes electroweak  interactions provides a
distribution rather than a simple $\delta$ function.   In this sense, provided assumptions specified above
 are fulfilled, our approach is
completely model independent.



\section{Quantitative computation}\label{sec:3}

We now discuss  in more technical terms the inclusion of EW
gauge boson emission through the evolution equations.
We start from a first principle definition of the energy spectrum for
emitted particles  and then we define the fragmentation functions as
statistical objects describing the probability of a particle to be
transformed into another with a certain momentum fraction.
The full evolution equations for the fragmentation functions, containing EW and QCD interactions, are
 analyzed. We provide  an expression that can be used to
 match the outcome of Monte Carlo codes adding EW corrections  at leading order ${\cal O}(\alpha_2)$. Our approach is similar in spirit to the one used, in a  different  context, in \cite{Barbot}, with important differences.

The relevant quantity for indirect signals of DM
is the energy spectrum $dN_f/dx$ of stable SM particles $f= \{e^+,e^-,\gamma,p,\bar{p},\nu,\bar{\nu},\bar d\}$ produced per
DM decay/annihilation, where
$x={2E_f}/{\sqrt{s}}$ ($0\le x\le 1$)
is the fraction of center of mass energy carried
by a stable particle  $f$  with energy $E_f$.
For clarity we will sometimes specify the formul\ae{} assuming the case of non-relativistic
DM annihilations, for which $\sqrt{s} =2M$ such that $x = E_f/M$; it is immediate
to obtain the corresponding formul\ae{} for DM decays, where $\sqrt{s}=M$.


\medskip

We assume that DM initially produces two primary back-to-back SM particles, and we consider all relevant cases:
\beq \begin{array}{rl}I=&\{ e^+_{L,R}e^-_{L,R},\  \mu^+_{L,R}\mu^-_{L,R}
,\ \tau^+_{L,R}\tau^-_{L,R}, \ \nu_e\bar\nu_e,\nu_\mu\bar\nu_\mu,\nu_\tau\bar\nu_\tau,\\
& q\bar q, \ c\bar c, \ b\bar b,\  t\bar t,\ \gamma\gamma,\ gg,
\ W^-_{T,L}W^+_{T,L} ,\  Z_{T,L}Z_{T,L}, \ hh\}\end{array}
,\eeq
where $q ={u,d,s}$ denotes a light quark; $h$ is the Higgs boson; $L$eft or $R$ight are the possible fermion polarizations,
and $T$ransverse or $L$ongitudinal are the possible polarizations of
massive vectors, that correspond to different EW interactions. Then, the
spectrum can be written as:
\beq
\frac{dN_f}{dx}
\equiv \frac{1}{\sigma_{{\rm DM~DM}\to I}}\frac{d\sigma_{{\rm DM~DM}\to I\to f+X}}{dx},
\qquad
f= \{e^+,e^-,\gamma,p,\bar{p},\nu,\bar{\nu},\bar d\},
\eeq
with a similar formula for the case of DM decay. The ``$X$'' in this
equation reminds of the inclusivity already discussed in Section 2.

\smallskip

In each one of the possible cases $I$, MonteCarlo generators like PYTHIA allow to compute the inclusive spectrum
${dN^{\rm MC}_{I\to f}}(M,x)/{dx}$ by generating
events starting from the pair $I$ of initial SM particles with back-to-back momentum and energy $E=M$,
and letting the MC to simulate the subsequent particle-physics evolution, taking into account
decays of SM particles and their hadronization, as well as QCD radiation and (partially) QED radiation.

Then, the spectra for a generic DM model that produce combinations of the two-body states $I$ can be
obtained combining the various channels:
\beq \frac{dN_f}{dx} = \sum_I {\rm BR}_I \frac{dN^{\rm MC}_{I\to f}}{dx} .
\eeq
In some DM models, primary multi-body states can be important: one can obtain the final spectra without running a dedicated
MC code by computing the model-dependent
energy spectra $D_I(z)$ of each primary pair $I$ (each one has energy $E = zM$ with $0\le z\le 1$)
and convoluting them with the basic MC spectra:
\beq \label{eq:Nbody}
\frac{dN_{f}}{d\ln x} (M,x)=\sum_J \int_x^1 dz \, D_{J}(z)\;\frac{dN^{\rm MC}_{J\to f}}{d\ln x}\left(zM,\frac{x}{z}\right).\eeq
Notice that we combine particle-antiparticle pairs because we assume that they have the same spectra, which is true
whenever the cosmological DM abundance does not carry a CP asymmetry.
Otherwise, hadronization can be significantly affected and dedicated MC runs would be necessary.
The indices  $I,J = p + \bar p$ denotes a primary particle $p$ together with
its anti-particle $\bar p$, with the same
energy spectrum. Factors of two are  such that for complex
particles one has $dN_{p}/dz = dN_{\bar p}/dz =D_I$, while for real particles
 (the $Z$, the $\gamma$, the Higgs $h$) one
has $dN_{{\rm DM}\to p}/dz = 2\;D_I$.

\subsection{Including EW corrections}

We now come back to the basic case of DM that annihilates or decays in one primary channel $I$ and
discuss how to achieve the goal of this paper: obtaining a set of basic functions $dN_{I\to f}/dx$ that take into account
EW radiation, replacing the functions $dN_{I\to f}^{\rm MC}/dx$ computed via MonteCarlo simulations.
EW radiation is a model-independent subset of the higher order corrections discussed above, and gives rise to specific
spectra of initial SM particles, such that its effect can be included in the primary basic spectra by a formula similar to eq.\eq{Nbody}:
\beq \label{eq:master}
\frac{dN_{I\to f}}{d\ln x} (M,x)=\sum_J \int_x^1 dz \,;D^{\rm EW}_{I\to J}(z)\;\frac{dN^{\rm MC}_{J\to f}}{d\ln x}\left(zM,\frac{x}{z}\right),\eeq
where $D^{\rm EW}_{I\to J}(z)$ is the EW $I\to J$ {\em EW parton distribution}: the $J$ spectrum produced by initial $I$. Our normalization
is such that we have
the uniform normalization $D^{\rm EW}_{I\to J}(z) = \delta_{IJ}\delta(1-z)$ at tree level for both real and complex particles.
Some comments are in order:

i) When including higher order effects, one must avoid overcounting and take into account that MC codes already include some particularly relevant higher-order effects: showering produced by strong (QCD) and
electromagnetic (QED) interactions, up to details.\footnote{We must include via eq.\eq{master} only all
those effects not included in MC codes.  Existing MC codes have their own peculiarities, e.g.\ {\sc Phytia}
automatically includes $\gamma$ radiation from charged particles but not from the $W^\pm$.   Of course,  an alternative more precise approach, that we do not purse, would be implementing the missing EW radiation effects into some existing MC code.}

ii) For initial particles that do have strong interactions, eq.\eq{master} misses the interplay between EW and QCD radiation;
this limitation is not a problem because, as expected, in such cases EW radiation will
turn out to be subdominant with respect to QCD radiation.

iii) For initial particles that do not have strong interactions, eq.\eq{master} holds at leading order in the weak couplings:
first they must do an EW splitting, and next one can add QCD splittings neglecting EW radiation.
We emphasize an important different between e.g.\ a $Z\to q\bar q$ splitting and the same $Z\to q \bar q$ decay: in the decay
the invariant mass of the $q\bar{q}$ pair is equal to the $Z$ mass (such that $Z\to t\bar t$ is forbidden by the heaviness of the top $t$);
in the splitting the invariant mass can be much higher, and we approximate it as $zM$.
This higher invariant mass strongly affects the subsequent QCD radiation from quarks, which is more abundant in the splitting case, leading to a higher multiplicity of $\bar p$ and $\gamma$.
 
In Appendix A we give a detailed discussion
of the interplay between EW and QCD radiation and of
the level of approximation introduced by using eq. (\ref{eq:master}).

\bigskip


\subsection{Computing the EW parton distributions}
We define $D^{\rm EW}_{I\to J}(z,\mu^2)$ as the probability for a given parent
 particle $I$ with virtuality of the order of $\mu$ to become a particle
 $J$ with a fraction $z$ of the parent particle's energy mediated by EW interactions.
 At large virtuality, they take the tree level values:
 \begin{equation}\label{boundaries}
D_{I\to J}^{\rm EW}(z,\mu^2=s)=\delta_{IJ}\;\delta(1-z);
\end{equation}
At low virtuality $\mu^2\sim M_W^2$, they are the functions we need: $D_{I\to J}^{\rm EW}(z) = D_{I\to J}^{\rm EW}(z,\mu^2=M_W^2) $.
The evolution in the virtuality is described by  integro-differential equations, that involve a set of kernels\footnote{In this work we indicate with $P(x,\mu^2) $ the {\sl unintegrated} kernels, while the splitting functions $P(x)$, obtained   by integrating in $\mu^2$, depend only on the energy fraction $x$; a list of the relevant splitting functions is given in Table \ref{tab:splitting}.  } $P^{EW}_{I\to J}(x,\mu^2)$ that have been derived in \cite{EvEq}:
\begin{equation}\label{EEqns}
\frac{\partial D^{\rm EW}_{I\to J}(z,\mu^2)}{\partial \ln\mu^2}=-\frac{\alpha_2}{2\pi}\;
\sum_k \;\int_x^1\frac{dy}{y}\; P^{\rm EW}_{I\to K}(y,\mu^2)\;D^{\rm EW}_{K\to J}(z/y,\mu^2).
\end{equation}
Since we work at leading order in the EW couplings, eq.~(\ref{EEqns})
with the boundary conditions of eq.~(\ref{boundaries}) is solved by:
\begin{equation}\label{soluzioni}
D_{I\to J}^{\rm EW}(z)=\delta_{IJ}\delta(1-z)+\frac{\alpha_2}{2\pi}\int_{M_W^2}^{s}\frac{d\mu^{2}}{\mu^{2}}\,
P_{I\to J}^{\rm EW}(z,\mu^{2}).
\end{equation}
Differently from QED and QCD, the EW kernels feature infrared singular terms proportional to $\ln\mu^2$, so that the solutions (\ref{soluzioni}) also include double logs beside the customary single logs of collinear origin:
\beq\label{eq:coll}
D^{\rm EW}_{I\to J}(z)=D_2(z)\,\ln^2\frac{M}{M_W}
+D_1(z)\,\ln\frac{M}{M_W}
+D_0(z).
\eeq
Our goal is to include the model-independent logarithmically enhanced terms.
Electroweak radiation from the initial DM state is of course model-dependent: since DM is non-relativistic this effect
only contributes to the non-enhanced terms $D_0$, that we neglect. Notice however that
for our purposes we need to include terms of the form $( \ln x)/x$ that are relevant in the region $x\to 0$; this is discussed in detail in subsection \ref{massive}.

\begin{table}[t]
$$
\begin{array}{rlrll}
 \multicolumn{2}{c}{\hbox{splitting $1\to x +x'$}} & \multicolumn{3}{c}{\hbox{splitting function: real and virtual}}\\ \hline
F_{0,M}\to& F_{0,M}+V_M & \displaystyle P_{F\to F} =& \displaystyle \frac{1+x^2}{1-x} L(1-x)&
P_{F\to F}^{\rm vir} =\displaystyle \frac{3\ell}{2}-\frac{\ell^2}{2}
\\[1mm]
F_{0,M}\to& V_M+F_{0,M}  & \displaystyle   P_{F\to V} =&\displaystyle  \frac{1+(1-x)^2}{x} L(x) &
P_{F\to V}^{\rm vir} =\displaystyle \frac{3\ell}{2}-\frac{\ell^2}{2}
\\[1mm]
V\to& F+\bar F & \displaystyle P_{V\to F} =&  \displaystyle [x^2+(1-x)^2]\ell &
P_{V\to F}^{\rm vir} = \displaystyle -\frac{2\ell}{3}
\\[4mm]
S_M\to &S_M+V_M& \displaystyle   P_{S\to S} = &\displaystyle2\frac{x}{1-x}L(1-x) &
P_{S\to S}^{\rm vir} =\displaystyle 2\ell-\frac{\ell^2}{2}\\[1mm]
S_M\to &V_M+S_M& \displaystyle   P_{S\to V} =& \displaystyle 2\frac{1-x}{x}  L(x)&
P_{S\to V}^{\rm vir} =\displaystyle 2\ell-\frac{\ell^2}{2} \\[2mm]
V\to &S+S' & \displaystyle P_{V\to S} =&  \displaystyle  x(1-x)\ell &
P_{V\to S}^{\rm vir} =\displaystyle -\frac{\ell}{6}\\[3mm]
V_M\to &V_M+V_M &  \displaystyle  P_{V\to V} =&
\multicolumn{2}{l}{\displaystyle 2 \bigg[\frac{x}{1-x}L(1-x)+\frac{1-x}{x} L(x)+ x(1-x)\ell\bigg]}\\[3mm]
V_M\to &V_M+V_0 &  \displaystyle  P'_{V\to V} =&
\multicolumn{2}{l}{\displaystyle 2 \bigg[\frac{x}{1-x}\ell +\frac{1-x}{x}L(x)+ x(1-x)\ell\bigg]}\\[3mm]
V_M\to &V_0+V_M &  \displaystyle  P_{V\to \gamma} =&
\multicolumn{2}{l}{\displaystyle 2 \bigg[\frac{x}{1-x}L(1-x)+\frac{1-x}{x} \ell + x(1-x)\ell \bigg]_{\phantom{1}}}\\[2.5mm]
 F\to&F+S & \displaystyle   P_{F\to F}^{\rm Yuk} =&(1-x)\ell & P_{F\to F}^{\rm Yuk,vir} =\displaystyle -\frac{\ell}{2} \\[2mm]
 F\to&S+F & \displaystyle   P_{F\to S}^{\rm Yuk} =&x\ell & P_{F\to S}^{\rm Yuk,vir} =\displaystyle -\frac{\ell}{2} \\[2mm]
  S\to&F+F & \displaystyle   P_{S\to F}^{\rm Yuk} =&\ell & P_{S\to F}^{\rm Yuk,vir} =\displaystyle -{\ell}
\end{array}
$$
\caption{\em Generalized splitting functions for massive partons.
$V$ denotes a vector, $F$ a fermion and $S$ a scalar;
$V_M$ denotes a vector with mass $M_V$, $V_0$ a massless vector,  etc.
The function $L(x)$ is defined in eq.~(\ref{Lexact}) and $\ell=\ln s/M_V^2$.
\label{tab:splitting}}
\end{table}

\subsection{Splitting functions}
The leading-order parton distributions $D_{I\to J}(z)$ can  be computed
by using the partonic splitting functions $P$ summing over all possible SM splittings~\cite{EvEq};
the relevant splitting functions  are here collected in table~\ref{tab:splitting}.

A concrete simple example allows to clarify the procedure and the normalization factors: we consider
DM producing an initial generic $F$ermion-anti$F$ermion pair with $F\bar {F}$ invariant mass $\sqrt{s}\gg m_F$.
We assume that $F$ has charge $q_F$ under a generic vector $V$ with mass $M_V\ll \sqrt{s}$ and gauge coupling $\alpha$.
(In the SM the vector could e.g.\ be a $Z$ and the fermion a neutrino).
The splitting process $F\to FV$ gives rise to:
\beq D_{F\to F}(z)=  \delta(1-z) \left[1 +  \frac{\alpha \; q_F^2}{2\pi}  P^{\rm vir}_{F\to F}\right] +
\frac{\alpha \; q_F^2}{2\pi} P_{F\to F}(z) ,\qquad
D_{F\to V}(z)= \frac{\alpha\; q_F^2}{2\pi} P_{F\to V}(z). \eeq
By replacing $F\to S$ one obtains the corresponding result for a pair for $S$calars, and so on.

The first term describes {\em virtual corrections} arising from one-loop diagrams, and the second term describes real emission corrections.
We define:
\beq \label{cwec} P_{I\to J}^{\rm vir} \equiv  -\int_0^1 dz~P_{I\to J}(z), \eeq
for any $I$ and $J$; e.g. from $P_{V\to V}$ in table~1 we have:
\begin{equation}\label{eg}
P_{V\to V}^{\rm vir}=\frac{11}{3}\ell-\ell^2\qquad\hbox{where}\qquad \ell=\ln {s\over M_V^2}.
\end{equation}
While both virtual corrections, related to $P_{I\to J}^{\rm vir}$, and
real corrections, related to $P_{I\to J}$, are  listed in table~1,
the simple relation (\ref{cwec}) holds between them.
The relationship between real and virtual contributions is dictated by  the
unitarity of the theory (see e.g.~\cite{CF} for a more detailed discussion).
Intuitively, this just amounts to
say that when an $F$ radiates, it disappears from the initial state.

One can verify that the partonic distributions $D_{I\to J}$ satisfy a set of conservations laws
(with corresponding identities for the splitting functions $P$)
\begin{itemize}
\item The conservation of splitting probability:
\begin{equation}\label{splittingprobability}
    \frac{1}{2}\sum_{J} \int_0^1 dx ~D_{I\to J}^{\rm real}  + \int_0^1dx ~D_{I\to I}^{\rm vir}= 0 ,
\end{equation}
where the factor $1/2$ accounts for the fact that one particle splits in two.

\item The conservation of total momentum:
\beq \sum_{J} \int_0^1 dx ~x~D_{I\to J} = 1.\eeq

\item The conservation of electrical charge:
\beq \sum_{J} \int_0^1 dx ~Q_J~D_{I\to J} = Q_{I}.\eeq
\end{itemize}
By combining these splitting functions with the appropriate electroweak couplings (including the top quark Yukawa coupling)
we get the electroweak splittings among SM particles, described by the $D_{I\to J}$ functions, explicitly listed in
Appendix~\ref{frag}.  It is  a simple exercise to verify that conservation laws are satisfied.
For completeness, we also list the splittings involving
photons (to be dropped if already included by MC codes), assuming for simplicity
a photon with mass $M_W$; sending it
 to zero gives infrared divergences that can be regulated and dealt with
   using well known techniques that we do not need to discuss here.


\subsection{Splitting functions for massive partons\label{massive}}
We see from table~1 that $P_{I\to J}^{\rm vir}$ is explicitly given by linear or quadratic polynomials in $\ln s/M_V^2$: the vector mass
provides an infra-red regulator.  This is unlike in the standard application of partonic techniques, where all partons are massless and some infra-red regularization is needed.

In this subsection we briefly describe how we generalized partonic functions to massive partons (such as the $W,Z,t$ in the SM),
and why particle masses modify the kinematics affecting  the log-enhanced terms, as encoded in a non trivial universal function $L$.

Let us consider the splitting process $i\to f+f'$ of a particle $i$ into massive partons $f,f'$: the corresponding splitting functions can
be non zero only in the kinematically allowed ranges:
\beq  \frac{m_f}{E_i}< x<1- \frac{m_{f'}}{E_i},\eeq
where $x=E_f/E_i$, and similarly for $x'=1-x = E_{f'}/E_i$.
The modification due to nonzero masses
in the kinematical range is power suppressed and thereby mostly negligible;
however the small forbidden regions at small $x,x'$ are relevant when {\em vectors} are emitted, because the corresponding splitting functions have
$1/x$ and $1/x' =1/(1-x)$ divergences, see table~1.
Thereby small parton masses {\em must} strongly affect such splitting functions, replacing the usual $1/x$ and $1/x'$ divergent terms with
new functions that vanish at the kinematical boundary.

\begin{figure}
\begin{center}
$$\includegraphics[width=0.55\textwidth]{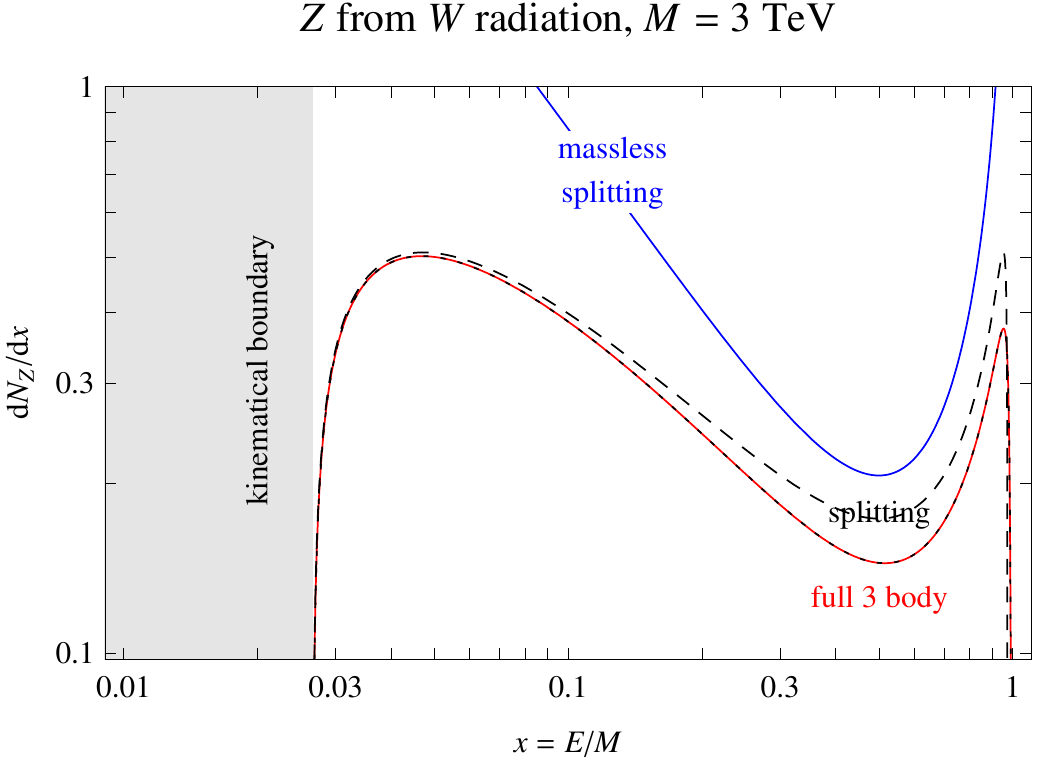}$$
\caption{\em {\bf Splitting function between massive vectors}.
Blue curve: naive result.  Dashed curve: our result.
Red curve: full 3-body result in the Minimal Dark Matter model.\label{fig:confronto1}}
\end{center}
\end{figure}

\medskip

We therefore need to correct the expressions for the splitting functions that are singular in the limits $x\to 0$ and $x'\to 0$ (soft $f$ and soft $f'$).
These singularities arises when soft vectors are emitted, such that the
correct expressions can be obtained using the {\em eikonal approximation},
which dictates the amplitude for emitting a soft vector.
Since we stop at leading order, we can precisely define the splitting functions in terms of the energy spectra of the particles produced in
three body scatterings: we just need to integrate such squared amplitude
over the massive phase space in the region that produces the leading
singularities.
Broadly speaking, what happens is that the upper and lower limits of
integration
that appear in (\ref{soluzioni}) are modified and become $x$-dependent;
computations and checks are performed in Appendix~\ref{integrazionekt}.
The resulting splitting functions for massive partons are listed in table~\ref{tab:splitting}, and, for $x\to0$,
they do not depend on spin, e.g.\ $P_{S\to V}\simeq P_{F\to V}\simeq P_{V\to V}$, as dictated by the eikonal amplitude.
They contain the universal kinematical function $L(x)$,
that replaces the usual $\ell=\ln s/M_V^2$ that holds for massive partons:
\begin{equation}\label{Lexact}
L(x)=
\ln\frac{sx^2}{4M_V^2}+2\,\ln\left(
1+\sqrt{1-\frac{4M_V^2}{sx^2}}
\right).
\end{equation}
This function indeed vanishes below the kinematical threshold ($x<2M_V/\sqrt{s}$) and reduces to $\ln sx^2/M_V^2$ well above it.
This means that {\em small parton masses, apart from providing kinematical thresholds,
give rise to extra $\ln x$ factors with respect to the standard case of massless partons}, which become numerically relevant
at small $x,x'$.

This was not noticed before, and Fig. \ref{fig:confronto1} exemplifies its relevance.
The dashed curve in Fig. \ref{fig:confronto1} shows the splitting function between massive vectors
(e.g.\ relevant for $Z$ radiation from $W^\pm$): it significantly differs from the massless splitting function (upper curve)
even away from the kinematical boundaries, and it closely agrees with the full result of a 3-body computation in a specific model,
which also includes not log-enhanced terms.  These terms are subleading in the present example, where  we here considered DM DM $\to W^+_T W^-_T$ annihilations with $M=3\TeV$.

\section{Results}

\begin{figure}[p]
\begin{center}
$$\includegraphics[width=0.45\textwidth]{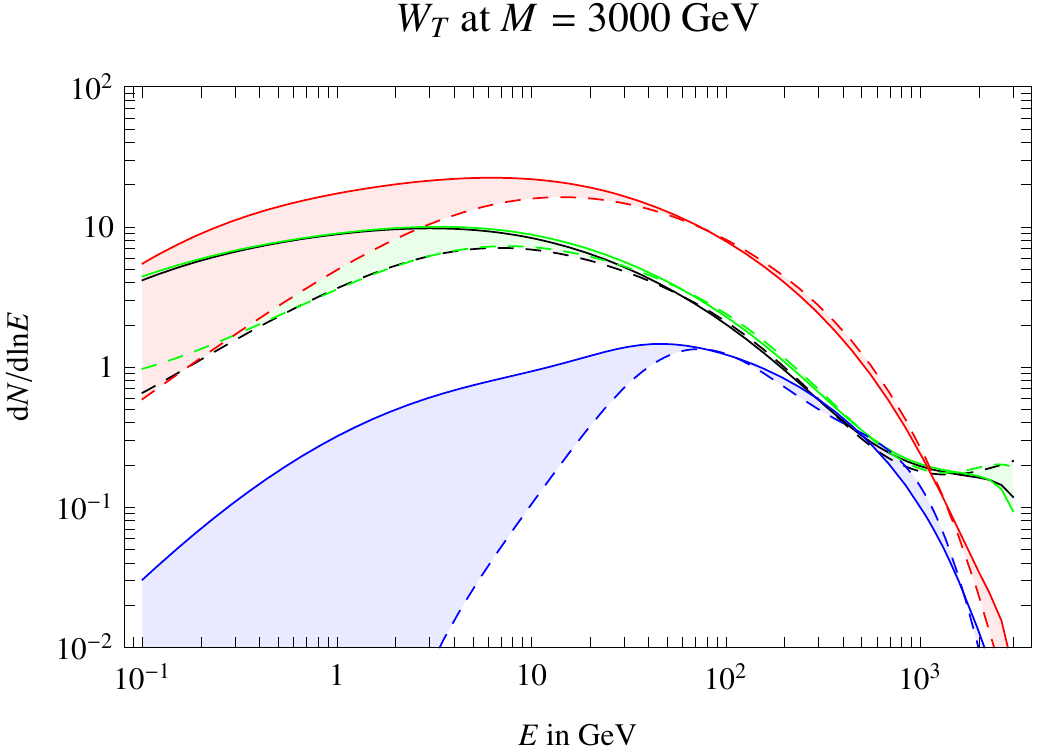}\qquad \includegraphics[width=0.45\textwidth]{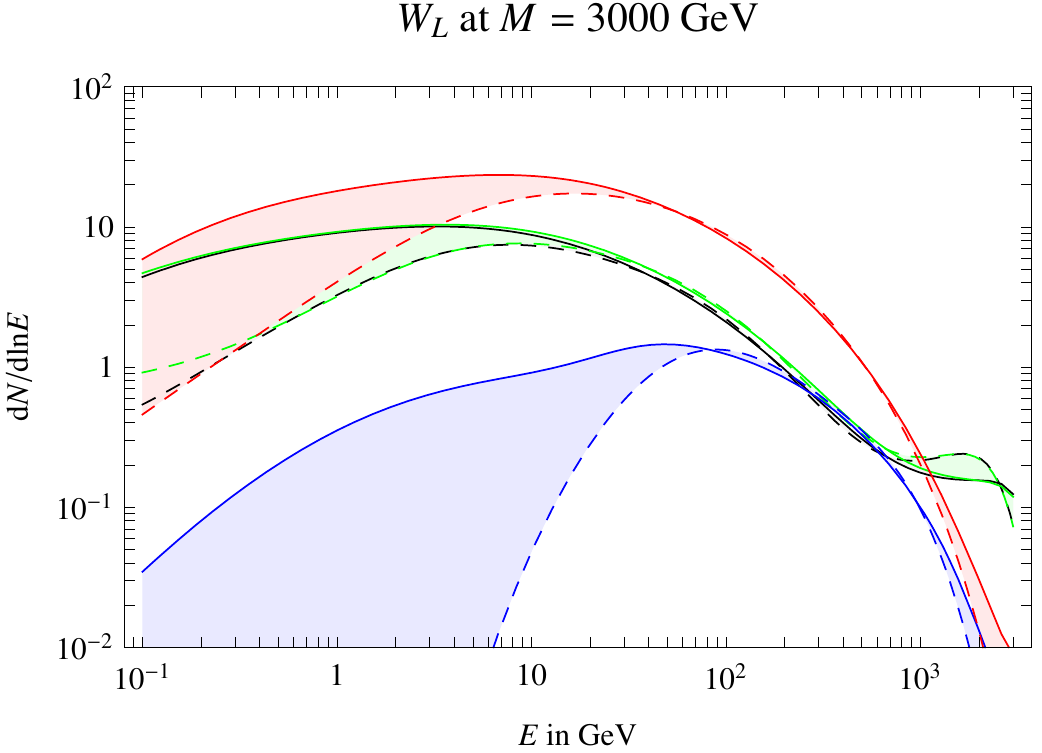}$$
$$\includegraphics[width=0.45\textwidth]{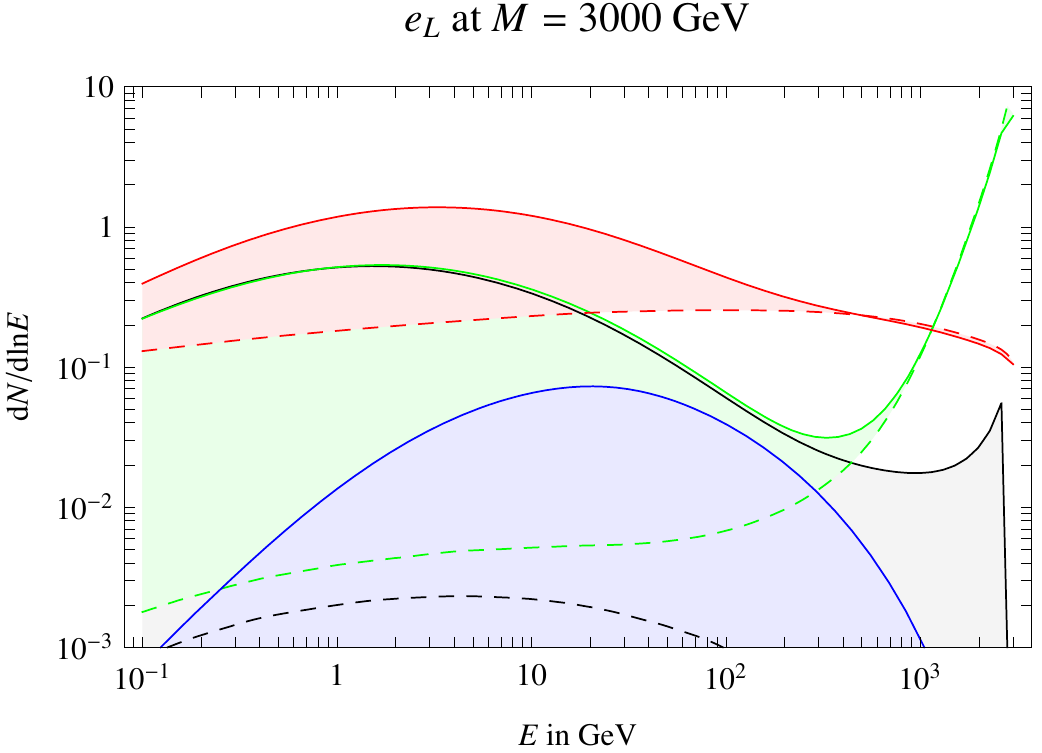}\qquad \includegraphics[width=0.45\textwidth]{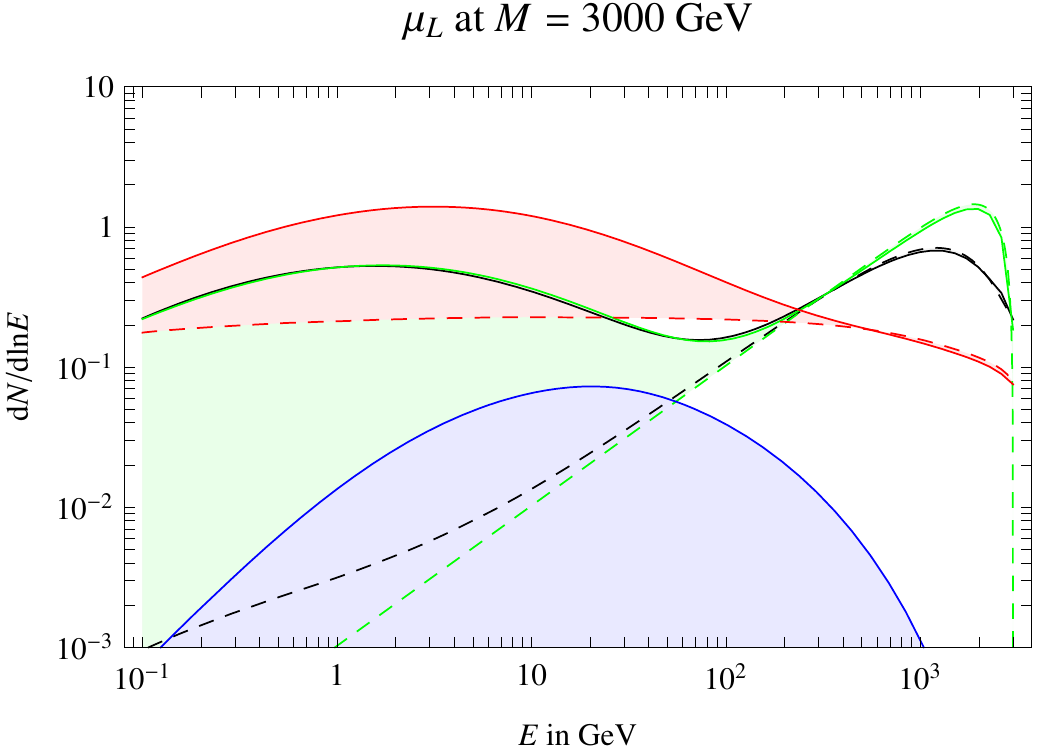}$$
$$\includegraphics[width=0.45\textwidth]{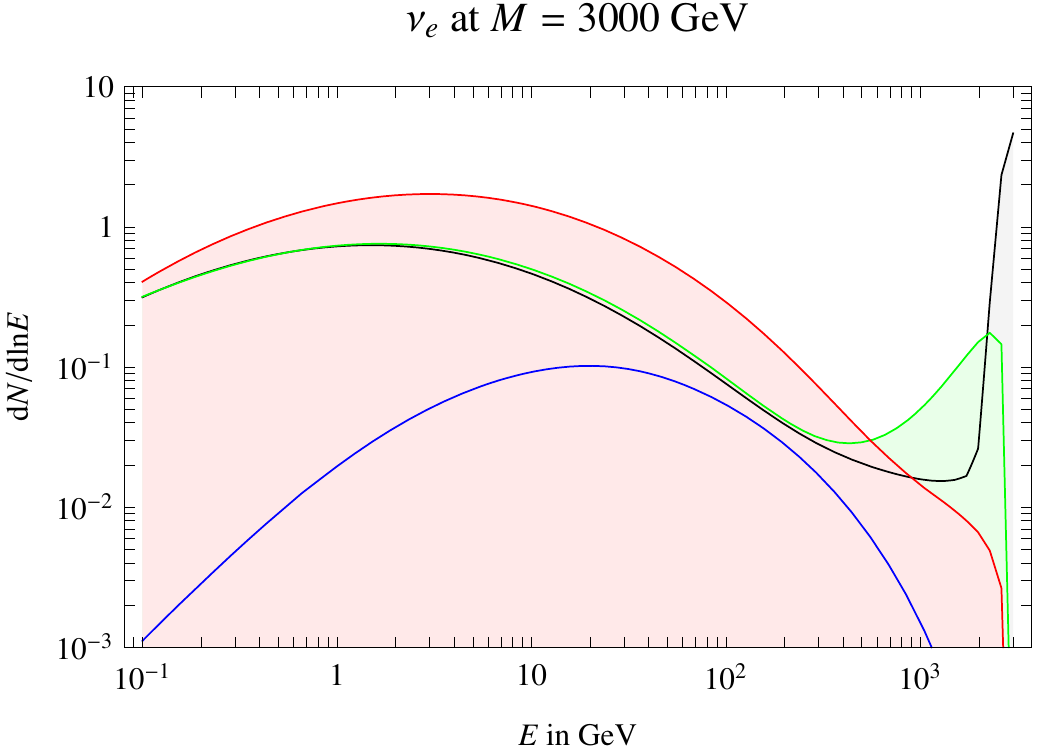}\qquad \includegraphics[width=0.45\textwidth]{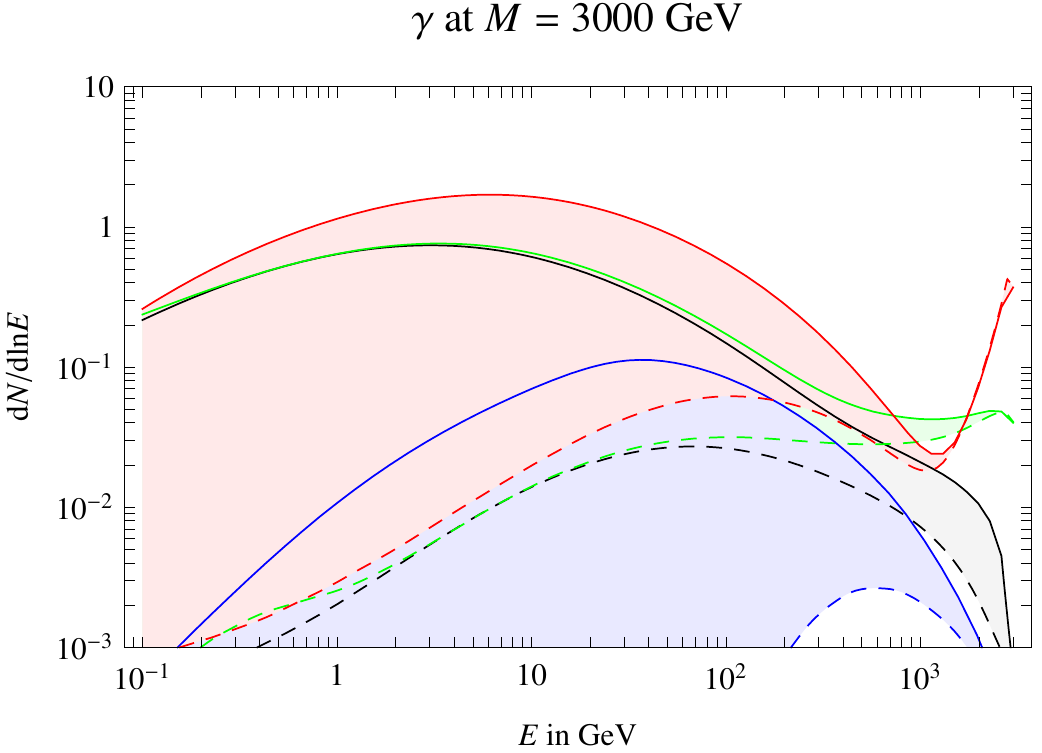}$$
\caption{\em Comparison between spectra with  (continuous lines) and without EW corrections (dashed).
We show the following final states: {\color{green}$e^+$ (green)}, {\color{blue}$\bar p$ (blue)}, {\color{red}$\gamma$ (red)}, $\nu = (\nu_e +\nu_\mu+\nu_\tau)/3$ (black).
\label{fig:CP}}
\end{center}
\end{figure}

\begin{figure}[t]
\begin{center}
$$\includegraphics[width=0.75\textwidth]{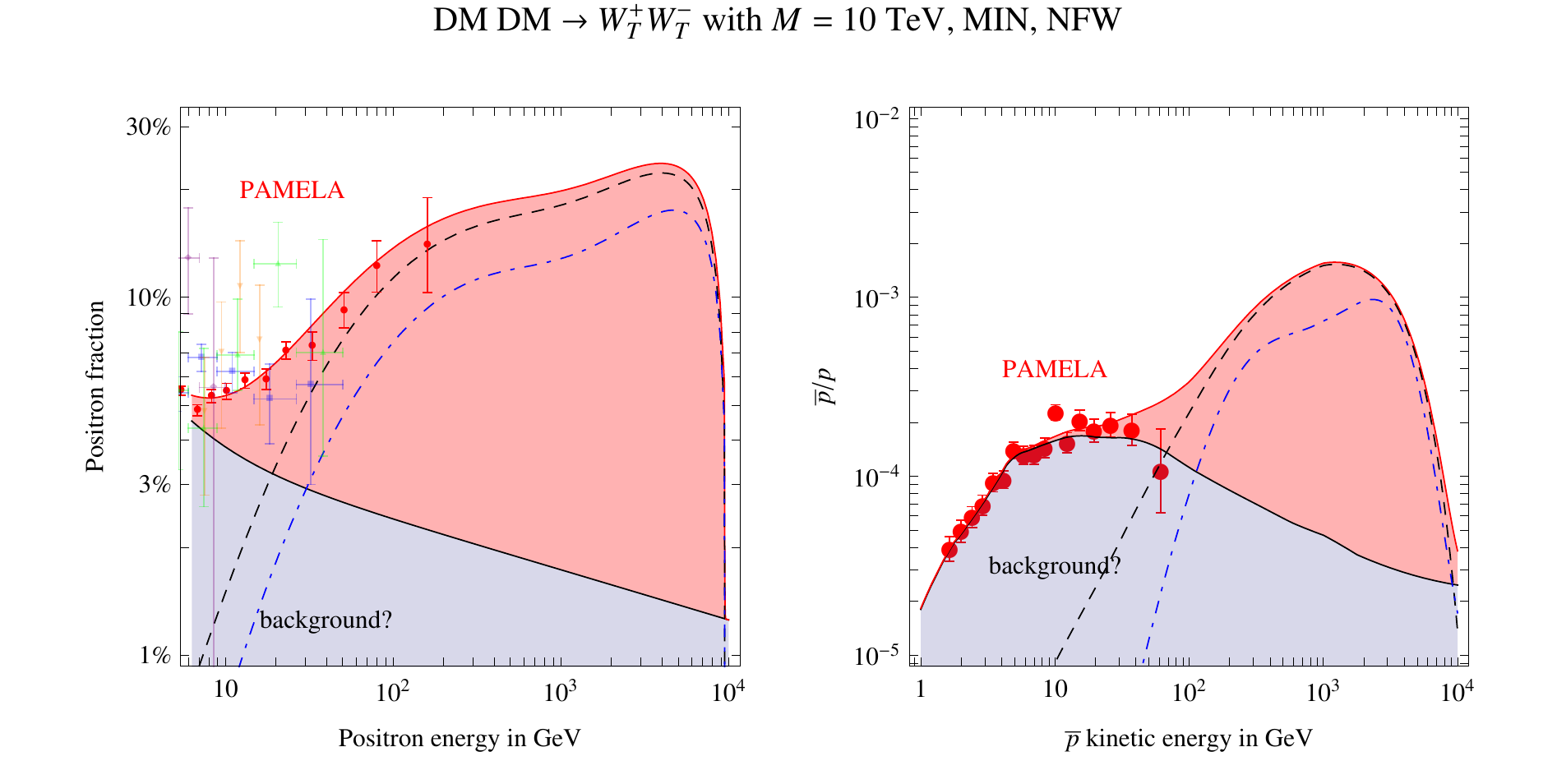}$$
$$\includegraphics[width=0.75\textwidth]{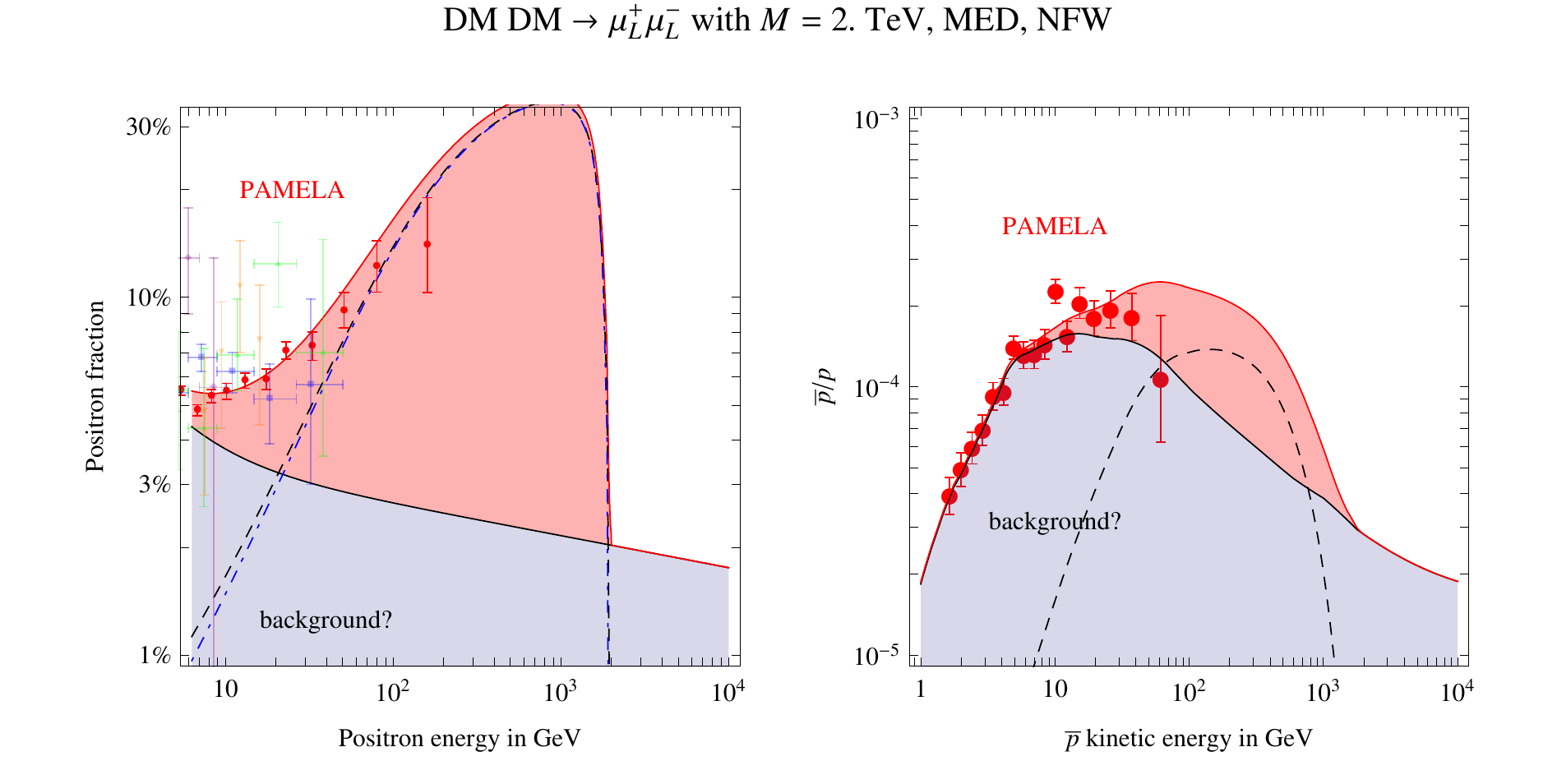}$$
\caption{\em DM signals in the $e^+$ (left) and $\bar p$ (right) fraction, with (dashed) and without (dot-dashed) electroweak corrections for two DM models that can fit the PAMELA $e^+$ excess: Minimal Dark Matter (upper) or a muonic channel
(lower). The gray area is the predicted astrophysical background and the red area is the prediction adding the full DM contribution.
\label{fig:PAM}}
\end{center}
\end{figure}

Our main results are the energy spectra of all stable final SM particles $f$ from any two-body DM non-relativistic annihilations or decays.
We will make the code freely available in~\cite{Cirelli}, and show here some examples.

In table~\ref{tab:N} we show the total number of $e^+$, $\bar p$ and $\gamma$, including EW radiation, for a few values of the DM mass $M$.
Without EW radiation, increasing $M$ would just boost all particles, such that $dN/dx$ and consequently $N$ does not depend on $M$.
With EW radiation, the total number of particles increases with the DM mass: the typical effect is about a factor of $2$ going from $0.3$ to 3 TeV.
Of course, such dependence on $M$ is similar to the one already present in the quark channels (bottom row), due to QCD radiation.

\begin{table}
$$
\begin{array}{l|ccc|ccc|ccc}
& \multicolumn{3}{|c|}{N_e}&\multicolumn{3}{c|}{N_\gamma,~ x>10^{-5}}&\multicolumn{3}{c}{N_p}\\
\hbox{DM mass in TeV} &0.3&1&3&0.3&1&3&0.3&1&3\\ \hline
\DM\,\DM\to e_L^-e_L^+& 1.07 & 1.47 & 2.02 & 1.09 & 2.08 & 3.39 & 0.0148 & 0.0574 & 0.118 \\
\DM\,\DM\to  \mu_L^-\mu^+_L &1.17 & 1.56 & 2.11 & 1.07 & 2.06 & 3.37 & 0.0147 & 0.0572 & 0.118 \\
\DM\,\DM\to \tau_L^-\tau_L^+& 1.48 & 1.86 & 2.40 & 3.15 & 4.09 & 5.34 & 0.0147 & 0.0572 & 0.118 \\
\DM\,\DM\to W_T^-W_T^+& 15.1 & 18.9 & 24.8 & 31.3 & 39.8 & 53.2 & 1.50 & 1.92 & 2.62 \\
\DM\,\DM\to W_L^-W_L^+& 15.5 & 19.6 & 25.8 & 32.3 & 41.6 & 55.9 & 1.52 & 1.94 & 2.66 \\
\DM\,\DM\to hh& 27.3 & 30.5 & 35.5 & 59.9 & 67.1 & 78.6 & 2.22 & 2.60 & 3.26 \\
\DM\,\DM\to q\bar{q}& 22.6 & 34.6 & 48.9 & 47.5 & 73.9 & 107. & 2.47 & 3.89 & 5.71
\end{array}$$
\caption{\em Total number of $e^+$, $\gamma$, $\bar p$ for a few DM annihilation channels.
\label{tab:N}}
\end{table}

Coming to the energy spectra, these new particles appear at low energy.
In Fig. \ref{fig:CP} we consider annihilations of DM with mass $M=3\TeV$ and compare the spectra with (continuous curves) and without (dashed curves) EW corrections.

\begin{itemize}
\item
In the top row we consider DM annihilations into $W^+W^-$ with transverse (left) or longitudinal (right) polarization. The final spectra do not significantly depend on the polarization of the $W$, but this is an accident: a $T$ransverse
$W$ splits into all quarks as demanded by their weak interactions, while a $L$ongitudinal $W$ (being the charged Goldstone in the Higgs doublet) mainly splits into $t$ and $b$ quarks, as
demanded by the top quark Yukawa interaction.
In both cases EW splittings increase the total number of final particles ($e^+$, $\bar p$, $\gamma$, $\nu$) increases by a factor of almost 2.
The biggest effect is in the $\bar p$ spectra:  without EW corrections they are strongly suppressed below $E_p<m_p\cdot (M/M_W)\sim 100\GeV$,
because of the boost factor $M/M_W$ of the $W$. Adding EW corrections, a $W$ splits into quarks with lower energy, producing $\bar p$
also at lower energy.
This is important for interpretations of the PAMELA $e^+$ excess in terms of annihilations of very heavy DM particles,
e.g.\ $M=10\TeV$ as predicted by Minimal Dark Matter~\cite{MDM}. EW corrections make more difficult to avoid unseen effects in $\bar p$ by having all the DM effects above 100 GeV where we do not have data yet.
As illustrated in Fig. \ref{fig:PAM}, in view of EW corrections there is some tension with data, even assuming the MIN model of
diffusion of charged particles in the galaxy~\cite{minmedmax} (which gives
the minimal amount of $\bar p$ compatibly with cosmic rays data and theory).

\item
We do not plot our result for DM annihilations into the Higgs boson, because its mass and decay modes are not yet known.
The Higgs channel is similarly affected by EW corrections as the $W_L$ and $Z_L$ channels (because they are the Goldstone components of the Higgs doublet),  and the main effect is again the $h\to t$ splitting induced by the top Yukawa coupling.

\item In the middle row of Fig. \ref{fig:CP} we consider DM annihilations into charged leptons: $e_L^-e_L^+$ (left) and $\mu_L^-\mu_L^+$ (right).
The spectra are significantly affected by EW corrections, because leptons split into $W,Z$ bosons, finally producing
quarks, and consequently a copious tail of $e^+,\gamma$, $\bar p$ at lower energy.
The main new qualitative feature is the appearance of $\bar p$ from leptons, and in Fig. \ref{fig:PAM} (lower row) we show that DM annihilations into $\mu_L^+\mu_L^-$ with $M=2\TeV$
(a scenario motivated by the PAMELA $e^+$ and FERMI $e^++e^-$ anomalies) also gives a possibly detectable excess of $\bar p$.
We here assumed the favored MED model of propagation of charged particles in the Milky Way~\cite{minmedmax};
$\bar p$ can be suppressed down to a
negligible by considering the MIN model and/or DM annihilations into $\mu_R^+\mu_R^-$.
Indeed EW effects are more significant for left-handed leptons which have $\SU(2)_L$ interactions than for right-handed leptons which only have U(1)$_Y$ interactions.

\item The bottom left panel of Fig. \ref{fig:CP} shows that, in view of EW interactions, DM annihilations into neutrinos also induce a significant spectrum of
$\gamma$ and some $e^+$, $\bar p$.

\item The bottom right panel of Fig. \ref{fig:CP} shows how EW corrections affect DM annihilations into $\gamma\gamma$: an hypothetical $\gamma$ line at high energy $E_\gamma=M\sim\hbox{few TeV}$ must be accompanied by a comparable flux of  $\gamma$ with lower energy $E_\gamma \sim (10-100)\GeV$, where we have more data.\footnote{This effect, already partly present in QED, seems to be
not implemented into MonteCarlo codes.}
\item
Finally, DM annihilations and decays into quarks are negligibly affected, so that we do not show our results.
\end{itemize}
The same results hold for DM decays.  EW corrections are irrelevant in models where DM decays or annihilates into hypothetical `dark' particles, lighter than the weak scale, that finally decay back to SM particles.

\medskip

There is a striking difference between the two examples of  $e^+$ fraction in Fig. \ref{fig:PAM}: EW corrections are relevant in the upper case
($W_T$ channel)
and negligible in the lower case ($\mu_L$ channel).
At first sight, this is surprising, because Fig. \ref{fig:CP} shows that in both cases EW corrections induce significant low energy tails to the $e^+$ spectra at production.
The qualitative difference is due to $e^+$ energy losses in the Milky Way that also gives a tail of $e^+$ at low energy:
we explain how to understand which effect is dominant.
The $e^+$ flux can be approximatively computed neglecting galactic diffusion and taking into account only energy losses, with the result
\beq
\frac{dn_{e^-}}{dE}=\frac{dn_{e^+}}{dE}=  \frac{3m_e^2}{4\sigma_T u_\oplus}
 \frac{1}{E^2} \frac{\sigma v}{2} \left(\frac{\rho_\oplus}{M}\right)^2
 \int_{E}^{M}  dE'~\frac{dN_{e}}{dE'}
\label{eq:fluxpositrons}
\eeq
where $\sigma_T$ is the Thompson cross section, $u_\oplus$ is the energy density in radiation and magnetic fields and $\rho_\oplus$ is the DM density, both at the location of the solar system.
This means that the $e^+$ flux at a given energy $E<M$ is proportional to the number of $e^+$ produced with energy between $E$ and $M$.
In the case of DM annihilations into leptonic channels, the tail of $e^\pm$ at lower energy produced by EW radiation contains a number or
$e^\pm$ at most comparable to the amount of $e^\pm$ already present at $E\sim M$: therefore EW corrections negligibly affect the
positron fraction.  This is not the case for DM annihilations into $W^\pm$, where instead the tail at low energy is the dominant component
of the total $e^\pm$ number, such that EW corrections more significantly affect the $e^+$ fraction.


%



\section{Conclusions}
\noindent
In this paper we computed the energy spectrum of SM particles stemming out
from DM annihilation/decay. We have shown that EW corrections have a
relevant impact on such spectra when the mass $M$ of the DM particles is
larger than the EW scale $M_W$. Soft EW boson emission is  enhanced
in the collinear and infrared regime and this leads to
$\ln^2 M^2/M_W^2$ enhancement factors. The result of the inclusion of EW
corrections is that all stable particles are present
in the final spectrum, independently of the primary annihilation/decay
channel. For instance, even if the DM particles annihilate/decay into
light neutrinos, EW corrections cause the  appearance of    hadrons and
photons in the very final spectrum.
The inclusion of EW corrections is therefore  an essential ingredient in
order to have a physical picture of the
correlated energy spectra of final stable particles. Our quantitative
results may be inferred from Fig. \ref{fig:CP} where
the energy spectra $dN/dE$ of stable SM particles are presented with and
without EW corrections. These spectra are the necessary
ingredients to predict the flux for indirect searches once the effect of
diffusion and galactic energy loss are included. As a rule of thumb we may
say that EW corrections are important in determining the final flux of
stable particles whenever

\begin{itemize}
\item the final flux of stable particles is dominated by the low energy
tail of the  $dN/dE$. One example is the case of
DM annihilation/decay into gauge bosons and $e^\pm$ final states (see fig.\fig{CP}).
This point becomes more relevant in the present experimental situation, where we mostly observe
particles below 100 GeV, possibly much below the DM mass.

\item the final flux of stable particles  is absent  when EW corrections
are not taken into account. One example is the case of
DM annihilation/decay into leptons and antiprotons $\bar p$ final states (see fig.\fig{PAM}).
This point is important also for neutrino fluxes
from DM annihilations in the sun or in the earth, because all SM particles, even those that loose energy in matter before decaying
into neutrinos, can radiate a $W$ or a $Z$ that promptly decays into neutrinos.

\end{itemize}
EW corrections may also significantly affect
the fluxes of particles generated from heavy gravitino decays in 
supersymmetric  theories. The computation of these energy fluxes is crucial in studying 
the  dissociation of the light element abundances generated during a 
period primordial nucleosynthesis.

We expect that EW radiative corrections have a minor effect on the freeze-out cosmological DM abundance,
because it is determined by just the total non-relativistic annihilation DM cross section.
In the case of energy spectra instead, as explained in this paper, the low energy tails can be enhanced by orders of magnitude,
while the high energy part of the spectrum is mildly depleted.
The net effect on the total number of final particles typically is an enhancement by a factor of 2.
We conclude that, when DM is around or above the TeV scale, one must take into account radiative EW corrections.

\smallskip

We computed EW corrections at leading order. Although we cannot give a sharp answer, we argued that it is not necessary to
resum higher order EW corrections as long as DM is not too heavy:
$\alpha_2 \ln^2(M^2/M_W^2)/2\pi \ll  1$ at $M\ll 100\TeV$.
Indeed, higher order corrections  are expected to give small effects for total cross sections at the TeV scale: a one loop effect of the order of 30\% means that one expects higher order effects to be at the 1\% level. 
It is more difficult to guess how low-energy tails might be affected without performing a dedicated computation,
which could be done implementing EW corrections in a MonteCarlo: we provided splitting functions for massive partons and
all ingredients.

%



\small

 \paragraph{Acknowledgments:}
 D.C. during this work was  partially supported by the
EU FP6 Marie Curie Research \& Training Network "UniverseNet"
(MRTN-CT-2006-035863).
P.C. wishes to thank Luigi Lanzolla for many useful contacts and
discussions, without which his work would have been much more difficult. We also thank G.F. Giudice
for useful conversations.
We used the high-statistics Pythia spectra computed on the Baltic Grid by Mario Kadastik, to appear in~\cite{Cirelli};
this work was supported by the ESF Grant 8499 and by the MTT8 project.

\numberwithin{equation}{section}
\appendix

\section{Evolution Equations }\label{eqv}
At present the energy spectrum $dN^{\rm MC}_{I\to
  f}/dx
$ are computed
with MC generators like PYTHIA
 by generating
events starting from the pair $I$ of initial SM particles with back-to-back momentum and energy $E=M$,
and letting the MC to simulate the subsequent particle-physics evolution, taking into account
decays of SM particles and their hadronization, as well as QED and QCD radiation.
As it is evident the MC output are nothing else that
the full QCD+QED fragmentation functions $D^{\rm QCD+QED}_{I\to
  f}(x)$ that can be identified with
 $dN^{\rm MC}_{I\to f}/dx$.
   The fragmentation functions are related to a nonperturbative aspect
  of QCD, so that they
 cannot be precisely
calculated by theoretical methods at this stage. The situation
is similar to the determination of the PDFs, where
high-energy experimental data are used for their determination
instead of theoretical calculations.
  The $\mu^2$ evolution for the fragmentation functions is calculated
in the same way as the one for the PDFs by
using the timelike DGLAP  evolution equations.
   The splitting functions are the same in the
Leading Order (LO)  evolution of the PDFs; however, they are different
in the Next-to-Leading Order (NLO). Explicit forms of the splitting functions are
provided in \cite{EvEq}.
The evolution equations are then essentially the same as
the PDF case.

The correct determination of the energy spectrum $dN_{I\to
  f}/dx$ of the final stable particle $f$, obtained through the particle-physical evolution of the initial pair $I$ 
  of SM particles with back-to-back momentum and energy $E=M$, needs the solution of a full set of evolution equations, including both strong and electroweak interactions.\\
In previous works the QCD DGLAP formalism has
been extended to EW interactions \cite{EvEq}.
The analysis of mass singularities in a spontaneously broken gauge theory like the electroweak
sector of the Standard Model has many interesting features. To begin with, initial
states like electrons and protons carry nonabelian (isospin) charges; this feature causes the
very existence of double logs, {\it i.e.} the lack of cancellations of virtual corrections with real
emission in inclusive observables \cite{BN}. Secondly, initial states that are mass eigenstates
are not necessarily gauge eigenstates; this causes some interesting mixing phenomena analyzed
in \cite{generic}.

On a general setting, the mathematical structure
of the full system of EW and  QCD
evolution is provided by
 the following  set of integral differential equations
 with kernels $P^{\rm QCD/EW}$ for  different gauge boson exchange
 (we omit for the moment the explicit QED evolution equations included in the EW part of the
SM and the various  index parametrizing the flavour and quantum numbers): \footnote{
The precise index (flavour) structure is given in the Appendix \ref{frag}, the
generic index structure is of the form
$\mu^2\frac{\partial }{\partial \mu^2}D_{I\to J}=\frac{\alpha}{2
  \pi}\sum_K\;
D_{I \to K}\otimes P_{K\to J}$ while
the $\otimes$-operator means $(f\otimes g)(x)\equiv f(z) \otimes g(x/z)=
  \int^1_x dz/zf(z)g\left(x/z
  \right)=\int^1_0dz\int^1_0dy f(y)g(z)\delta(x-zy)$. }
 \begin{eqnarray}
\label{ewqcd}
\mu^2\frac{\partial}{\partial \mu^2}D(x,\mu^2)&=&
\frac{\alpha_s}{2 \pi}\; D \otimes P^{\rm QCD}\;\theta(\Lambda < \mu < M)
+\frac{\alpha_2}{2 \pi}\;\;D \otimes  P^{\rm EW}\;\theta(M_W< \mu <M)\nonumber\\
&=&D \otimes \left (\frac{\alpha_s}{2 \pi}\; P^{\rm QCD}+\frac{\alpha_2}{2 \pi}\;  P^{\rm EW}  \right)
\;\theta (M_W< \mu <M)\nonumber\\
&+&\frac{\alpha_s}{2 \pi}\;D \otimes  P^{\rm QCD}\;\theta (\Lambda_{\rm QCD}< \mu <M_W),
\end{eqnarray}
where we have made explicit the  running range: for the EW interactions  from $M$
to $M_W$ where  the gauge boson mass is freezing the running
 and for QCD corrections from $M$ to
 $\Lambda_{\rm QCD}$ where   the non-perturbative effects  generate an effective cutoff to gluon exchanges.
 In practice,  from $\Lambda_{\rm QCD}$ to $M_W$ the running is purely dictated by QCD (remember  that QED interactions are  not shown for simplicity, in this case the running scale for photons  is until the electron mass
 $m_e$) while the EW-QCD interplay starts only above the $M_W$ scale.
Note also that  forgetting the underlying
index structure can bring to the wrong conclusion that, simply due to
the fact that $\alpha_s >\alpha_2$, the EW part is just a correction to
the QCD dominant part.
This would be a mistake as  in many interesting annihilation channels (like all the leptonic ones or
the $W^{\pm}$, $Z$ or $h$)  the QCD part is simple zero.

 A numerical solution to the full (EW+QCD) problem is of course out of reach.
 Nevertheless, we can find some reasonable  approximation taking advantage of the fact that   we can simulate the  pure QCD evolution also in the non perturbative regime with MC codes
  and  that EW theory is in the perturbative regime.
    Technically specking, the evolution equations are  Schr\"{o}dinger-like equations with a time dependent Hamiltonian,  where time is replaced by the $\mu^2$ variable and the Hamiltonian by the
    $ P$- kernel.  A formal solution can be parametrized  with the evolution operator:
  \begin{eqnarray}
   \label{general}
  D(x,\mu^2_1,\mu^2_2)&\equiv&
  U(z,\mu_1^2,\mu^2_2)\otimes \;{\bf I}\;\delta\left(1-\frac{x}{z}\right)\nonumber\\
  & =&\left(
  {\cal P}_{\mu^2}\;
  e^{\int^{\mu_1^2}_{\mu_2^2}\frac{d \mu^2}{\mu^2}\left (\frac{\alpha_s}{2 \pi}\; P^{\rm QCD}+\frac{\alpha_2}{2 \pi}\;  P^{\rm EW}  \right)}\right)\;{\otimes}\;{\bf I},
  \end{eqnarray}
  where ${\cal P}_{\mu^2}$ is the $\mu^2$-ordering operator and ${\bf I}$ the identity in the flavour space.
  Due to the linearity of the Eqs.  (\ref{ewqcd}) \footnote{It might be useful to remember the property
  $  D(z,M^2,M_W^2)\otimes D(x/z,M_W^2,
  \Lambda_{\rm QCD}^2)=D(x,M^2,\Lambda_{\rm QCD}^2) $. }
  we can then  formally write the full solution as:
  \beq \label{geneq}
  D(x,M^2,\Lambda^2_{\rm QCD} )=
  U(z,M^2,M_W^2)\;\otimes \; D^{\rm QCD}\left(\frac{x}{z},M_W^2,\Lambda^2_{QCD} \right),
  \eeq
  where we have  separated the running from $M$ to $M_W$ inside
the evolution operator $U$ (that can be perturbatively expanded as soon as $\alpha_s\ln M^2/M_W^2$ and   $\alpha_2\ln^2 M^2/M_W^2$ are smaller than unity)  and the purely QCD piece
$D^{\rm QCD}(x,M_W^2,\Lambda^2_{\rm QCD} )$ encoding  also the non-perturbative low energy physics.
In order to keep  under control  further  simplifications we need also to  know  the matrix flavour structure. We display it  under the form of a  simplified four-dimensional space spanned by  $l$=leptons, $W=(W^\pm,Z,\gamma)$, $q $=quarks and  $g$=gluons,  for  the EW and QCD kernels, $ P^{\rm QCD}$ and $ P^{\rm EW}$:
{\small
\beq\label{ker}
 P^{\rm QCD}=\left(
\begin{array}{cccc}
 0 & 0 & 0 & 0 \\
 0 & 0 & 0 & 0 \\
 0 & 0 &  P_{{qq}}^{{\rm QCD}} &  P_{{qg}}^{{\rm QCD}} \\
 0 & 0 &  P_{{gq}}^{{\rm QCD}} & P_{{gg}}^{{\rm QCD}}
\end{array}
\right),
\;\; P^{\rm EW}=
\left(
\begin{array}{cccc}
 P_{{ll}} ^{{\rm EW}}& P_{{lW}}^{{\rm EW}} & 0 & 0 \\
 P_{{Wl}}^{{\rm EW}}& P_{{WW}}^{{\rm EW}} & P_{{Wq}}^{{\rm EW}} & 0 \\
 0 & P_{{qW}} ^{{\rm EW}}& P_{{qq}}^{{\rm EW}} & 0 \\
 0 & 0 & 0 & 0
\end{array}
\right).
\eeq
}
The above   matrices do not commute; the  EW and QCD sectors are connected through the channels
 $W\to q,\;q\to W$ and $q\to q$, furthermore  the leptonic and hadronic sectors are connected
through the mixed $W\to q,\;l,\;\;l,\;q\to W$  channels.
Reasonable approximate solutions are related both to the possibility to expand perturbatively the
general solution (\ref{general}), and to the outcome from MC generators which take automatically into account the full
QCD plus QED evolution from  $M$ to $\Lambda_{\rm QCD}$ scales ($m_e$ for QED).

One way of proceeding is to define pure EW ($D^{\rm EW}$) and QCD ($D^{\rm QCD} $) fragmentation functions that evolve with their  respective kernels, see eq. (\ref{ker}),  for the energy range $M_W^2<\mu^2<M^2$:
\beq\label{partialewqcd}
\mu^2\frac{\partial}{\partial\mu^2}D^{\rm QCD}(x,\mu^2)=\frac{\alpha_s}{2 \pi}\;
D^{\rm QCD} \otimes  P^{\rm QCD}\quad{\rm and}\quad
\mu^2\frac{\partial}{\partial\mu^2}D^{\rm EW}(x,\mu^2)=\frac{\alpha_2}{2 \pi}\;
D^{\rm EW} \otimes P^{\rm EW},
\eeq
whose  formal solutions are:
 \beq
  D^{\rm EW/QCD}(x,M^2,M_W^2  )=
  P_{\mu^2}\;
  e^{\int^{M^2}_{M_W^2}\frac{d \mu^2}{\mu^2}\left (\frac{\alpha_{2/s}}{2 \pi}\; P^{\rm EW/QCD}\right)_{\otimes}}  \;{\bf I}.
  \eeq
  Then we introduce  a new factorized EW $\otimes$ QCD fragmentation function:
  \beq\label{newd}
\overline{D}(x,\mu^2)\equiv
( D^{\rm EW}\otimes D^{\rm QCD}
)(x,\mu^2)
 \qquad {\rm with}\qquad\theta(M_W<\mu<M).
\eeq
This is clearly not  a solution of the true evolution equations
 (\ref{ewqcd}) but can be a useful approximate solution.
In order to relate the true solution  $D$ of eq. (\ref{geneq}) with
 the new function $\overline{D}$ satisfying eq. (\ref{newd})),
we can use the fact that in the $M_W^2<\mu^2<M^2$ interval we are
in perturbative regime also for the QCD side.
Knowing that for  two generic non-commuting  operators ${\bf A}$ and ${\bf B}$:
\beq
e^{{\bf A}+{\bf B}}=\left(
{\bf I}-\frac{1}{2}[{\bf A},{\bf B}]+...\right)\;e^{\bf A}\;e^{\bf B},
\eeq
and identifying
   ${\bf A}=\int^{M^2}_{M_W^2}\frac{d \mu^2}{\mu^2}\left
(\frac{\alpha_{2}}{2 \pi}\;P^{\rm EW}\right)$ and ${\bf
  B}=\int^{M^2}_{M_W^2}\frac{d \mu^2}{\mu^2}\left (\frac{\alpha_{s}}{2
  \pi}\; P^{\rm QCD}\right)$,  we can approximate
 the evolution operator $U$ at any order in $\alpha_{s,2}$.
In particular at second order in ${\cal
  O}(\alpha_{s,2}^2,\;\alpha_s\alpha_2)$
we have:
\begin{eqnarray}
U(x,M^2,M_W^2)&=&
\left[
 \left( {\bf I}+\frac{\alpha_s \alpha_2}{8 \pi^2}\;\;
 \int^{M^2}_{M_W^2}\frac{d\mu^2}{\mu^2 }\int
 \frac{d\mu^2}{\mu^2 }\;\;[ P^{\rm EW}, P^{\rm QCD}]+\cdots\right)\right.\nonumber\\
&\otimes&\left.\;\underbrace{D^{\rm EW}\otimes D^{\rm QCD}}_{\overline{D}}\right](x,M^2,M_W^2),
\end{eqnarray}
where:
\beq
[   P^{\rm QCD}, P^{\rm EW}]_{\otimes}\equiv
  \left(
   P^{\rm QCD}\otimes  P^{\rm EW}-  P^{\rm EW}\otimes  P^{\rm QCD}\right),
  \eeq
  and the $\cdots$ stand for the fact that  there is an infinite
  series of
 commutators with coefficients of order
$\alpha_2^{m+1}\;\alpha_s^{n+1}$ with $m+n\geq 1$.
Starting from this expression we can   write the perturbative relation
between
  the exact solution $D$ and the present outcome of the MC codes $D^{\rm QCD}$:
\beq\label{fulleq}
D=
 \left( {\bf I}+\frac{\alpha_s \alpha_2}{8 \pi^2}\;\;
 \int^{M^2}_{M_W^2}\frac{d\mu^2}{\mu^2 }\int
 \frac{d\mu^2}{\mu^2 }\;\;[P^{\rm EW},P^{\rm QCD}]+...\right)
\;\otimes\;D^{EW}\otimes \frac{dN^{\rm MC}}{dx}.
\eeq%
The first order corrections to such a formula are obtained expanding also $D^{\rm EW}$ at one loop:
\beq\label{eweq}
D^{\rm EW}(x,\mu^2)=\delta(1-x)\;{\bf I }\;+\frac{\alpha_2}{2
  \pi}\; \int^{s}_{\mu^2} P^{\rm EW}(x,\mu'^2)\frac{d \mu'^2}{\mu'^2},
\eeq
where we have explicitly shown the arguments $x$ and $\mu$ of the matrix $ P^{\rm EW}$.\\
At order  ${\cal O}(\alpha_2\ln^2 M^2/M_W^2,\alpha_s\ln M^2/M_W^2)$  we find that
the  energy spectrum for the process $I\to f+X$ can be therefore
written as in eq. (\ref{eq:master}):
    \begin{eqnarray}
\frac{d N_{I\to f }}{dx}&=&\sum_J
\;\left(I_{IJ}\;+\frac{\alpha_2}{2
  \pi}\; \int^{s}_{M_W^2}  P^{\rm EW}_{I\to J}(x,\mu'^2)\frac{d
  \mu'^2}{\mu'^2} \right)\otimes D^{\rm QCD}_{J\to f}\left(\frac{x_f}{x_I},M^2,\Lambda^2\right)\nonumber\\
  &\equiv&\label{princ}
\sum_J\;
\left(I_{IJ}\;+\frac{\alpha_2}{2
  \pi}\; \int^{s}_{\mu^2} P^{\rm EW'}_{I\to J}(x,\mu'^2)\frac{d
  \mu'^2}{\mu'^2} \right)\otimes \frac{d N_{J\to f }^{\rm MC}}{dx},
\end{eqnarray}
where in the last passage, to be consistent,  we have written  ${\rm EW'}$ in order to
stress that only massive $W^{\pm}$ and $Z$ are included while QED is
already encoded in  the MC.

Through  this expression one can match the MC code with
the first order EW corrections.

\section{Eikonal approximation and the improved splitting functions}\label{integrazionekt}

The standard partonic approximation holds in QED and in QCD for the emission of soft massless gauge bosons (photons or gluons) from partons, showing the presence of universal logarithmical factors of collinear origin that multiply the usual splitting functions.
This approach can't be na\"{\i}vely applied to the electroweak case when a massive gauge boson, such as the $W$, is involved in the splitting process; considering for definiteness $i\to f+f^{\prime}$ and defying $x\equiv E_{f}/E_{i}$, in fact, the allowed kinematical range for this latter is:
\begin{equation}\label{kinrange}
\frac{m_f}{E_i}\leq x\leq 1-\frac{m_{f^{\prime}}}{E_i},
\end{equation}
where particle masses act as cut-off for the soft singularities at $x\to 0,1$. These boundary regions in (\ref{kinrange}) are therefore extremely important and the standard partonic approximation have to be improved, introducing extra $\ln x$ and $\ln(1-x)$ terms, well justified by the kinematical proprieties of the splitting process.

In this Appendix we derive our improved splitting functions for massive partons, following the logic outlined in section \ref{sec:3}.
In \ref{eikonal} we use the eikonal approximation, that describes the amplitudes with soft gauge boson emission.
We integrate it over the phase space using for it the Sudakov approximation in \ref{parametrizzazionedisudakov} 
and using the exact expression in \ref{paresatta}: the Sudakov parametrization, commonly used in literature, don't respect the boundaries in (\ref{kinrange}).
In \ref{CA} we introduce, through an explicit example, the collinear approximation and its propriety of factorization. 
Finally, in \ref{fullresult} we compare our results with those of a full full three body calculation (exact amplitude integrated over the exact phase space).

\subsection{The eikonal amplitude}\label{eikonal}
As well known, the spin of the emitting particles (scalar, fermion and vector) becomes irrelevant in the eikonal limit:
for definiteness, and without losing generality we here consider the real emission of a particle with momentum $k$ described through the three gauge boson vertex $3g$.

\begin{figure}
\begin{center}
\includegraphics[width=10 cm]{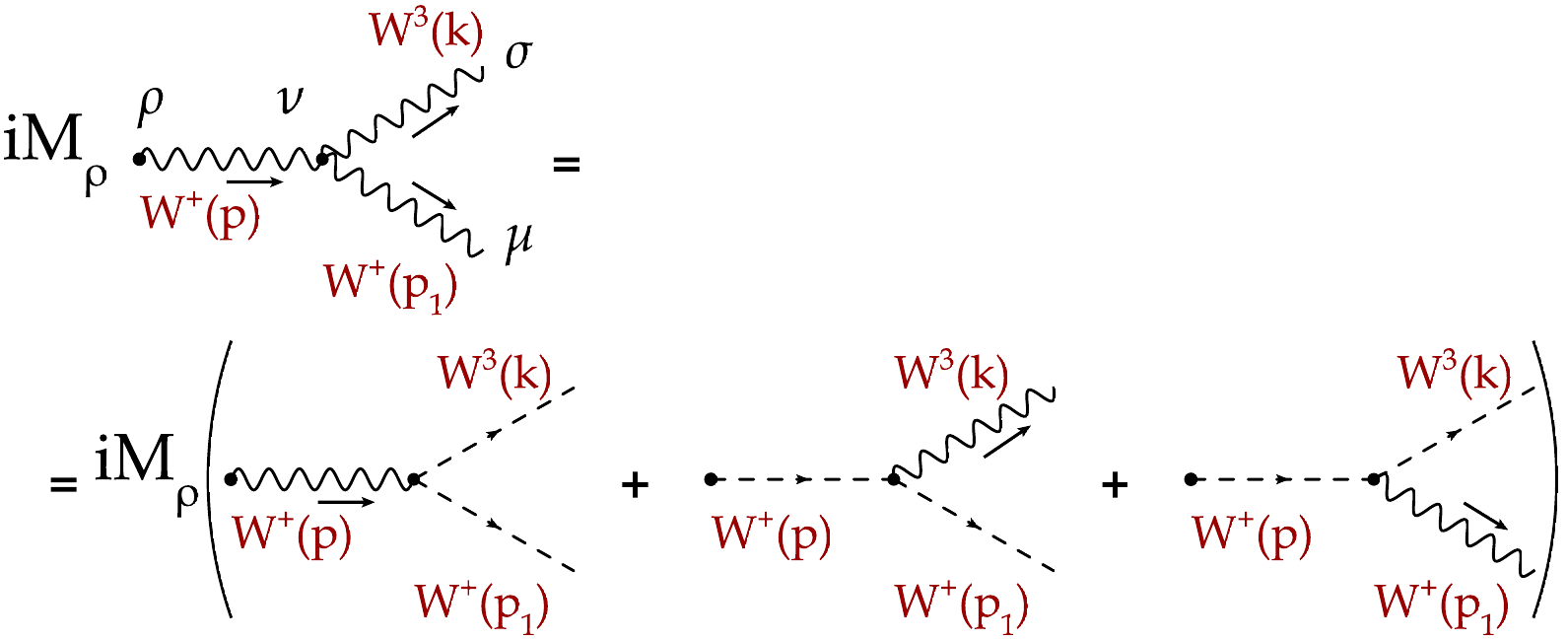}\\
\caption{{\em Soft gauge boson real emission from spin-1 particle. It can be read as the sum of three scalar currents. Considering the process $\sqrt{s}\to p_1+k+p_2$ we show explicitly only the bremsstrahlung contribution from the $p_1$ final leg.}}\label{3gsudakov}
\end{center}
\end{figure}
Considering Fig. \ref{3gsudakov} and using the conservation of momenta in the splitting vertex $p=p_1+k$ we can write:
\begin{equation}\label{3gfeynmansudakov}
i\mathcal{M}^{\rho}\left\{
\frac{-g}{2p_{1}\cdot k}\left[
\varepsilon^{*}(k)\cdot\varepsilon^{*}(p_{1})(k-p_{1})_{\rho}+\varepsilon^{*}_{\rho}(p_{1})2p_{1}\cdot \varepsilon^{*}(k)
-\varepsilon^{*}_{\rho}(k)2k\cdot \varepsilon^{*}(p_{1})
\right]
\right\},
\end{equation}
where we have indicated with $i\mathcal{M}^{\rho}$ the remaining part of the amplitude and taken for simplicity $k^2=0$.
Roughly speaking using the eikonal approximation it is possible to neglect the soft momenta in the numerator in front of the hard one and in this example we can discuss the two opposite situation in which either $p_1$ or $k$ are soft.

Considering Ward identities the first term in (\ref{3gfeynmansudakov}) vanishes in the case of transverse gauge bosons,
while for longitudinal degrees of freedom we refer the interested reader to \cite{Ciafaloni:2001vu} for an analysis of the electroweak symmetry breaking effects; therefore for our purpose we have:
\begin{equation}\label{3gfeynmansudakov2}
i\mathcal{M}^{\rho}\left\{
-\frac{g}{p_{1}\cdot k}\left[\varepsilon^{*}_{\rho}(p_{1})p_{1}\cdot \varepsilon^{*}(k)
-\varepsilon^{*}_{\rho}(k)k\cdot \varepsilon^{*}(p_{1})\right]
\right\},
\end{equation}
where:
\begin{itemize}

\item the first term reconstructs the hard scattering amplitude $i\mathcal{M}^{\rho}\varepsilon^{*}_{\rho}(p_{1})$ and it survives when $k$ is soft;

    \item the second term reconstructs the hard scattering amplitude $i\mathcal{M}^{\rho}\varepsilon^{*}_{\rho}(k)$ and it survives when $p_{1}$ is soft.
\end{itemize}

Squaring the amplitude (\ref{3gfeynmansudakov2}), we sum over polarizations using the axial gauge~\cite{Brock:1993sz}:
\begin{equation}\label{axial}
\sum\varepsilon^{\mu}(k)\varepsilon^{*\,\nu}(k)=-g^{\mu\nu}+\frac{k^{\mu}p_2^{\nu}+k^{\nu}p_{2}^{\mu}}{p_2\cdot k},
\end{equation}
and as a result the eikonal limit leads to the factorization of the process in the product of the hard cross section times an emission factor integrated over the allowed phase space of the soft particle:
\begin{eqnarray}
  \mathcal{I}(p_1) &=& g^2\frac{2k\cdot p_2}{(p_{1}\cdot k)(p_2\cdot p_1)}
  \frac{d^4p_1}{(2\pi)^3}\left.\delta(p_1^2)\right|_{p_1^0>0}
\,\,\,\,\,\,\,\,\,\,\,\,\,\,\,\,\,\,\mbox{$p_1$ soft},\label{Ip1}\\
\mathcal{I}(k) &=& g^2\frac{2p_1\cdot p_2}{(p_{1}\cdot k)(p_2\cdot k)}
  \frac{d^4k}{(2\pi)^3}\left.\delta(k^2)\right|_{k^0>0}
\,\,\,\,\,\,\,\,\,\,\,\,\,\,\,\,\,\,\,\,\mbox{$k$ soft}.\label{Ik}
\end{eqnarray}
The two integrals can be obtained through the exchange $p_1\leftrightarrow k$.

\subsection{The Sudakov parametrization}\label{parametrizzazionedisudakov}

We consider now the explicit evaluation of the eikonal integral in (\ref{Ip1}), and in order to perform this calculation we  choose a convenient parametrization of the external momenta;
fixing two basis vectors:
\begin{equation}\label{basisvector}
P=(E,0,0,E),\,\,\,\,\,\,\,\,\,\overline{P}=(E,0,0,-E),\,\,\,\,\,\,\mbox{with:}\,\,s=4E^2\simeq4M^2,
\end{equation}
the Sudakov parametrization consists in the following decomposition for the soft momentum $p_1$:
\begin{equation}\label{k}
p_1=xP+\overline{x}\overline{P}-k_{\perp}=\left(E(x+\overline{x}),-k_t,0,E(x-\overline{x})\right),
\end{equation}
where, without losing generality, we have taken the spatial component of $k_{\perp}$ along the $x$ direction.

In order to highlight in a simple way the logarithmical behavior of the eikonal integral we choose to work considering as first step the massless case
and approximating the two hard momenta  $k\simeq P$ and $p_2\simeq \overline{P}$, such that
\begin{equation}\label{misura}
d^4p_1=\frac{s\pi}{2} dk_t^2\,dx\,d\overline{x}.
\end{equation}
The eikonal integral takes the form:
\begin{equation}\label{eiko}
\mathcal{I}(p_1)=dk_t^2\,dx\,d\overline{x}\,\delta\left(sx\overline{x}-k_t^2\right)\,\frac{\alpha_2}{\pi}\,\frac{1}{x\overline{x}},
\end{equation}
and logarithmical singularities clearly arise in two opposite kinematical regions:
\begin{itemize}
\item $x\gg \overline{x}$: the soft gauge boson $p_1$ is emitted along the $k$ direction; integrating over $\overline{x}$ using $\overline{x}=k_t^2/sx$, the condition $x\gg \overline{x}$ becomes an upper bound for the transverse momentum $k_t^2\ll sx^2$ and therefore in terms of fragmentation functions in the $p_1$ soft limit $x\to 0$ we obtain:
\begin{equation}\label{primaintegrazionesudakov}
D_{x\to 0}=\int_{M_W^2}^{sx^2}dk_t^2\,\frac{\alpha_2}{2\pi}\,\frac{2}{x}\,\frac{1}{k_t^2}=
\frac{\alpha_2}{2\pi}\,\frac{2}{x}\,\ln\frac{sx^2}{M_W^2},
\end{equation}
that vanishes when $x=M_W/\sqrt{s}$.
\item $x\ll \overline{x}$: integrating over $x$ using the relation $x=k_t^2/s\overline{x}$,
the eikonal integral gives exactly the same logarithmical result previously discussed but in an opposite kinematical configuration since the soft gauge boson $p_1$ is emitted now along $p_2$ direction.
\end{itemize}

In order to clarify the consequences of the symmetry $p_1 \leftrightarrow k$ between the two eikonal integrals,
we can now discuss in more details the Sudakov parametrization in the region $x\gg \overline{x}$.
First we generalize eq. (\ref{k}) writing for $k$ and $p_2$:
\begin{eqnarray}\label{p1}
  k = zP+\overline{z}\overline{P}+k_{\perp}=\left(E(z+\overline{z}),k_t,0,E(z-\overline{z})\right), \quad
  p_2 = y\overline{P}=(yE,0,0,-yE),
\end{eqnarray}
and using the on-shell conditions in order to eliminate $\overline{z}$, $\overline{x}$ writing $\overline{z}=k_t^2/sz$,
$\overline{x}=k_t^2/sx$.
The conservation of energy and spatial momentum gives the following relations between the kinematical variables $x$, $y$, $z$:
\begin{eqnarray}
   y = 1-\frac{k_t^2}{4E^2z(1-z)}, \quad
 x = 1-z,
\end{eqnarray}
and we can therefore generalize the result for $k$ soft just considering the substitution
$p_1\to k\Longrightarrow x\to z= 1-x$, and as a consequence the kinematical end-points for the $x$ variable in the Sudakov parametrization are:
\begin{equation}\label{kinendpoints}
\frac{M_W}{\sqrt{s}}\leq x\leq 1-\frac{M_W}{\sqrt{s}}.
\end{equation}
 Comparing eq. (\ref{primaintegrazionesudakov}) with:
 \begin{equation}\label{general}
D=\frac{\alpha_2}{2\pi}\int\frac{dk_t^2}{k_t^2}P(x,k_t^2),
 \end{equation}
 where $P(x,k_t^2)$ is the usual unintegrated splitting function,
 we obtain its leading behavior in correspondence of the two kinematical limit in the Sudakov parametrization:
\begin{eqnarray}
   x\to \frac{M_W}{\sqrt{s}}:\hspace{1cm}P_{\rm Sud} &\sim& \frac{2}{x}\,\left.L(x)\right|_{\rm Sud},\label{sudakovfinale1} \\
   x\to 1-\frac{M_W}{\sqrt{s}}:\hspace{1cm}P_{\rm Sud} &\sim& \frac{2}{1-x}\,\left.L(1-x)\right|_{\rm Sud},\label{sudakovfinale2}
\end{eqnarray}
with:
\begin{equation}\label{Lxsudakov}
\left.L(x)\right|_{\rm Sud}=\ln\frac{sx^2}{M_W^2}.
\end{equation}

\subsubsection{Parton masses and the lower limit of integration}

The upper bound on the integration  over $k_t^2$ is dictated by the kinematical proprieties of the collinear emission, and can be studied even working in the massless case; parton masses, contrarily, affect in a relevant way the lower bound of integration.
To discuss this point, we need first to generalize the Sudakov parametrization in eq.~(\ref{p1}): it's straightforward to verify that one can take into account of the on-shell conditions $k^2=m_k^2$ and $p_1^2=m_1^2$ just redefining $k_t^2\to k_t^2+m_k^2$ for $k$ and $k_t^2\to k_t^2+m_1^2$ for $p_1$. Then from the propagator of the collinear emission $p\to p_1+k$ we have:
\begin{equation}\label{propagator}
\frac{1}{(p_1+k)^2-m_p^2}=\frac{x(1-x)}{k_t^2+m_1^2+x(m_k^2-m_1^2)-m_p^2x(1-x)}.
\end{equation}
Depending on which particles are massive, three different situations arise.
\begin{itemize}
\item[1.] Only the emitted vector is massive. This happens e.g.\ in EW interactions of the $W,Z$ vectors.
Assuming  $m_k^2=M_W^2$, $m_p^2=m_1^2=0$, eq. (\ref{propagator}) becomes:
 \begin{equation}\label{propagatorEW}
\frac{1}{(p_1+k)^2}=\frac{x(1-x)}{k_t^2+xM_W^2}\approx \frac{x(1-x)}{k_t^2}\,\vartheta(k_t^2-xM_W^2),
\end{equation}
where the latter passage holds in logarithmical accuracy. Integrating over $k_t^2$ we have for the $x\to 1$  singularity of the EW splitting function $P_{F\to F}$:
\begin{equation}\label{Pfflimit}
P_{F\to F}\sim \frac{2}{1-x}\ln\frac{s(1-x)^2}{xM_W^2}\simeq\frac{2}{1-x}\left.L(1-x)\right|_{\rm Sud},
\end{equation}
and therefore the lower $x$-dependence don't affect the soft limit $x\to 1$.

\item[2.]  Two particles are massive. This happens e.g.\ in the electromagnetic coupling of the $W$: $m_p^2=m_k^2=M_W^2$, $m_1^2=0$; eq. (\ref{propagator}) becomes:
 \begin{equation}\label{propagator2ggamma}
\frac{1}{(p_1+k)^2-M_W^2}\approx \frac{x(1-x)}{k_t^2}\,\vartheta(k_t^2-x^2M_W^2).
\end{equation}
Integrating over $k_t^2$ we have for the $x\to 0,1$  singularities of the splitting function $P_{V\to \gamma}$:
\begin{equation}\label{PVVlimit}
P_{V\to \gamma}\sim \frac{2}{1-x}\ln\frac{s(1-x)^2}{x^2M_W^2}+\frac{2}{x}\ln\frac{sx^2}{x^2M_W^2}\simeq
\frac{2}{1-x}\left.L(1-x)\right|_{\rm Sud}+\frac{2}{x}\ln\frac{s}{M_W^2}
,
\end{equation}
and the lower $x$-dependence affects the $x\to 0$ singularity for the soft photon.

\item[3.] All three particles are massive.  This happens in
the massive three gauge boson vertex, $m_p^2=m_k^2=m_1^2=M_W^2$; eq. (\ref{propagator}) becomes:
 \begin{equation}\label{propagator3g}
\frac{1}{(p_1+k)^2-M_W^2}\approx \frac{x(1-x)}{k_t^2}\,\vartheta[k_t^2-(1-x+x^2)M_W^2].
\end{equation}
Integrating over $k_t^2$ we have for the $x\to 0,1$  singularities of the splitting function $P_{V\to V}$:
\begin{equation}\label{PVVlimit}
P_{V\to V}\sim \frac{2}{1-x}\ln\frac{s(1-x)^2}{(1-x+x^2)M_W^2}+\frac{2}{x}\ln\frac{sx^2}{(1-x+x^2)M_W^2}\simeq
\frac{2}{1-x}\left.L(1-x)\right|_{\rm Sud}+\frac{2}{x}\left.L(x)\right|_{\rm Sud}
,
\end{equation}
and therefore the lower $x$-dependence don't affect the soft limits $x\to 0,1$.

\end{itemize}

\subsection{The exact parametrization}\label{paresatta}

The Sudakov parametrization, as we will discuss in \ref{fullresult} comparing our results with a full three body calculation, shows a bad behavior approaching $x \to 0$, namely when $p_1$ is soft.

This because in (\ref{k}) $x$ cannot be considered exactly the variable describing the fraction of energy of the particle after the splitting process respect to its initial value.
In order to correct this point, we need to introduce a different parametrization.
Referring to the eikonal integral (\ref{Ip1}) we write:
\begin{eqnarray}
  k &=& \left(zE,k_t,0,\sqrt{z^2E^2-k_t^2}\right), \\
  p_1 &=& \left(xE,-k_t,0,\sqrt{x^2E^2-k_t^2}\right), \\
  p_2 &=& \left(yE,0,0-yE\right);
\end{eqnarray}
and from energy and spatial momentum conservation we have:
\begin{equation}\label{conservazione}
\left\{\begin{array}{c}
         x+y+z=2, \\
         \sqrt{z^2E^2-k_t^2}+\sqrt{x^2E^2-k_t^2}-yE=0,
       \end{array}
\right.
\end{equation}
together with the conditions:
\begin{equation}\label{condizionicappat}
x^2E^2\geq k_t^2,\,\,\,\,\,\,\,\,\,\,\,z^2E^2\geq k_t^2,\,\,\,\,\,\,\,\,\,\,\,0\leq x,y,z\leq 1.
\end{equation}
In this parametrization $x$ can be considered exactly as the variable describing the fraction of energy, resolving the ambiguity noticed in the Sudakov parametrization.

The scalar products appearing in (\ref{Ip1}) can be explicitly rewritten as:
 \begin{eqnarray}
  p_1\cdot k &=& xzE^2+k_t^2-\sqrt{z^2E^2-k_t^2}\,\sqrt{x^2E^2-k_t^2},\label{p1dotk} \\
  p_2\cdot k &=& yE\left[zE+\sqrt{z^2E^2-k_t^2}\right],\label{p2dotk}\\
  p_1\cdot p_2 &=& yE\left[xE+\sqrt{x^2E^2-k_t^2}\right],\label{p1dotp2}
\end{eqnarray}
while for the phase space of the emitted soft particle we have:
\begin{equation}\label{phspace}
\frac{d^3\overrightarrow{p}_{1}}{(2\pi)^32p_1^0}=
\frac{dxdk_t^2}{16\pi^2}\,\frac{E}{\sqrt{x^2E^2-k_t^2}}.
\end{equation}
In order to simplify the integration we note that:
\begin{itemize}

\item in the soft limit $x\to 0$ from (\ref{conservazione}) it follows that $z\approx y\approx 1$,

\item since $x^2E^2\geq k_t^2$ and $0\leq x\leq 1$ it is possible to approximate $xE^2+k_t^2\approx xE^2$.

\end{itemize}
As a consequence the scalar products in (\ref{p1dotk},\ref{p2dotk},\ref{p1dotp2}) are simplified:
\begin{eqnarray}
  \left.p_1\cdot k\right|_{x\to 0} &=& E\left[xE-\sqrt{x^2E^2-k_t^2}\right], \\
  \left.p_2\cdot p_1\right|_{x\to 0} &=& E\left[xE+\sqrt{x^2E^2-k_t^2}\right],\\
  \left. k\cdot p_2\right|_{x\to 0} &=& 2E^2,
\end{eqnarray}
and the eikonal integral, considering (\ref{condizionicappat}) and introducing the mass $M_W$ as a physical cutoff for the $k_t^2\to 0$ singularity, reduces to:
\begin{equation}\label{integroesatta}
    D_{x\to 0}=\int_{M_W^2}^{sx^2/4}\frac{dk_t^2}{16\pi^2}\,\frac{g^2}{\sqrt{x^2E^2-k_t^2}}\,
\frac{4E}{k_t^2}=\frac{\alpha_2}{2\pi}\,\frac{2}{x}\,\left[
\ln\frac{sx^2}{4M_W^2}+2\ln\left(
1+\sqrt{1-\frac{4M_W^2}{sx^2}}
\right)
\right]
,
\end{equation}
that shows the behavior:
\begin{equation}\label{lb}
D_{x\to 0}\sim\frac{2}{x}\,\ln\frac{sx^2}{4M_W^2},
\end{equation}
when $sx^2/4\approx M_W^2$, vanishing correctly when $x=2M_W/\sqrt{s}$.

The symmetry propriety of the eikonal integral allow us to generalize this result for $k$ soft just considering the substitution $x\to z\approx 1-x$ and therefore the kinematical end-points on the $x$ variable for the exact parametrization are:
\begin{equation}\label{kinendpointsexact}
\frac{2M_W}{\sqrt{s}}\leq x\leq 1-\frac{2M_W}{\sqrt{s}}.
\end{equation}
As a conclusion we obtain the leading behavior of integrated splitting functions in correspondence of the two kinematical limit in the exact parametrization:
\begin{eqnarray}
  x\to \frac{2M_W}{\sqrt{s}}:\hspace{0.5cm}P_{\rm exact} &\sim& \frac{2}{x}\,L(x)
  ,\label{exactfinale1} \\
 x\to 1-\frac{2M_W}{\sqrt{s}}:\hspace{0.5cm}P_{\rm exact} &\sim& \frac{2}{1-x}\,L(1-x),\label{exactfinale2}
\end{eqnarray}
with $L(x)$ given in eq.~(\ref{Lexact}).


\subsection{The Collinear Approximation}\label{CA}
The eikonal approximation allows to highlight the singular behavior of the improved splitting functions in the soft regions $x\to 0,1$, as shown e.g. in Eqs. (\ref{exactfinale1},\ref{exactfinale2}). In order to extract the entire structure of the splitting functions and to show the factorization proprieties of our model independent approach, we need to go one step further, introducing the CA;
we discuss now its main features, having in mind an illustrative explicit example.
Following \cite{CF} we add to the SM Lagrangian a vector boson $Z^{\prime}$ with mass $M\gg M_W$ belonging to an extra ${\rm U}(1)'$ gauge symmetry and singlet under $\rm{SU}(3)_{C}\otimes \rm{SU}(2)_L\otimes {\rm U}(1)_Y$. In order to simplify our discussion, let us suppose that the $Z^{\prime}$ couples only with left electron and neutrino
\begin{equation}\label{couplingZ'}
\mathcal{L}_{\rm int}=f_L Z^{\prime}_{\mu}\overline{L}\gamma^{\mu}L,\hspace{2 cm}\mbox{with:}\;\;L=(\nu_L,e_L)^T.
\end{equation}
At tree level we have two possible leptonic decay channel $\Gamma_{ee}\equiv \Gamma_2(Z^{\prime}\to e_L^+e_L^-)$ and $\Gamma_{\nu\nu}\equiv \Gamma_2(Z^{\prime}\to \nu_L\overline{\nu_L})$, related through total isospin conservation
$\Gamma_{\nu\nu}=\Gamma_{ee}\equiv \Gamma_{B}$.
Considering the process $Z^{\prime}\to \nu_L \overline{\nu_L}$,
labeling
with $P=(E,0,0,E)$ and $\overline{P}=(E,0,0,-E)$ the two back-to-back
 momenta of the two neutrinos (with $E=M/2$), and indicating with
 $\Gamma_{2}(Z^{\prime}\to \nu_L \overline{\nu_L})$ the corresponding two-body decay width
for the amplitude squared we have  at Born level:
\begin{equation}\label{bornlevel}
\left|\mathcal{M}_{\rm
  Born}\right|^2=f_L^2Tr\left(\slashed{\overline{P}}
\gamma^{\mu}\slashed{P}\gamma^{\nu}\right)\varepsilon_{\mu}^{*}(Q)
\varepsilon_{\nu}(Q).
\end{equation}
We calculate now the effect of adding one weak gauge boson emission,
focusing on the three-body decay width $\Gamma_3$ related to the
process $Z^{\prime}(Q)\to \nu_L(p_1)\overline{\nu_L}(p_2)Z_T(k)$ and using the CA.

The key point of this approximation is the following: in the high energy regime $M\gg M_W$ the leading contributions to the three-body decay are produced by the region of phase space where the emitted boson is collinear either to the final fermion or
to the final antifermion, and in this region the three-body decay width is factorized with
respect to the two-body one.

Introducing:
\begin{equation}\label{3body}
d\Gamma_3=\frac{1}{2M}\left|\mathcal{M}_3\right|^2(2\pi)^4\delta(Q-p_1-p_2-k)\frac{d^3\overrightarrow{p}_1}{(2\pi)^3 2p_1^0}
\frac{d^3\overrightarrow{p}_2}{(2\pi)^3 2p_2^0}\frac{d^3\overrightarrow{k}}{(2\pi)^3 2k^0},
\end{equation}
it's possible to show in a simple way how the CA works both considering the factorization of the amplitude squared and of the phase space related to the final state.

\begin{figure}[!htb!]
\begin{center}
  \includegraphics[width=7 cm]{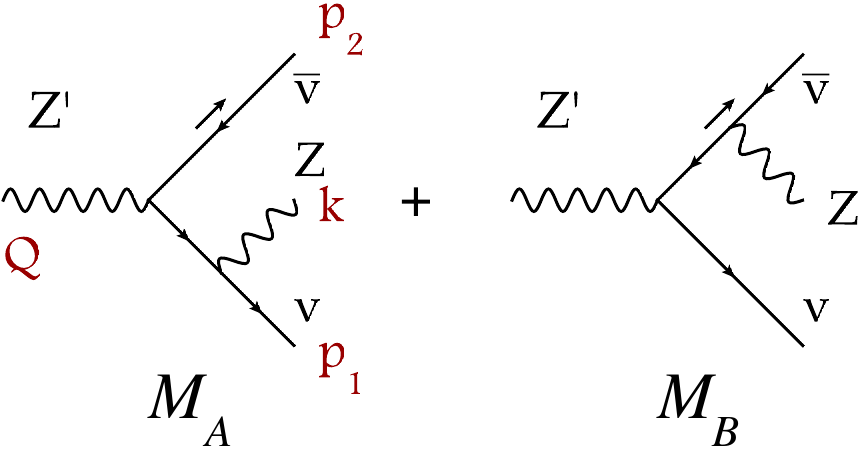}\\
  \caption{\emph{\rm Feynman diagrams involved in the calculation of the amplitude $\mathcal{M}_3=\mathcal{M}_A+\mathcal{M}_B$ of the decay process $Z^{\prime}(Q)\to \nu_L(p_1)\overline{\nu_L}(p_2)Z(k)$.}}\label{figuradiesempio}
  \end{center}
\end{figure}

Considering for definiteness the case in which the weak gauge boson $k$ is emitted along $p_1$ direction, we depict in Fig. \ref{figuradiesempio} the two Feynman diagrams involved in the computation of the amplitude $\mathcal{M}_3$.

Concerning the factorization of the amplitude squared we have, working for simplicity in the massless limit:
\begin{eqnarray}
  i\mathcal{M}_A &=& \frac{g}{2c_W}\,\frac{1}{2p_1\cdot k}\,\varepsilon_{\mu}^{*}(k)\left[\overline{u_L}(p_1)
  \gamma^{\mu}(\slashed{p}_1+\slashed{k})\Upsilon v_L(p_2)\right], \\
  i\mathcal{M}_B &=&  \frac{g}{2c_W}\,\frac{1}{2p_2\cdot k}\,\varepsilon_{\mu}^{*}(k)\left[\overline{u_L}(p_1)
 \Upsilon(\slashed{p}_2+\slashed{k})\gamma^{\mu}v_L(p_2)\right],
\end{eqnarray}
where $\Upsilon\equiv if_L\gamma^{\mu}\epsilon_{\mu}(Q)$;
therefore when the gauge boson is emitted along the $p_1$ direction $p_1\cdot k\to 0$, and the first diagrams diverges while the second one is finite; as a consequence squaring the amplitude $\mathcal{M}_3=\mathcal{M}_A+\mathcal{M}_B$ and using Feynman gauge divergences appear in the square $|\mathcal{M}_A|^2$ and in the interference term $\mathcal{M}_A\mathcal{M}_B^*+
\mathcal{M}_A^*\mathcal{M}_B$. Using the Sudakov parametrization \ref{parametrizzazionedisudakov} for the external momenta
it's straightforward to verify that:
\begin{equation}\label{MAsquared}
|\mathcal{M}_A|^2=\frac{g^2}{2c_W^2}\frac{1}{2p_1\cdot k}Tr\left(\Upsilon \slashed{p}_2 \Upsilon^* \slashed{k}\right);
\end{equation}
now we use the only technical point of CA; writing:
\begin{equation}\label{tecpoint}
2p_1\cdot k = \frac{k_t^2}{x(1-x)},
\end{equation}
we simply observe that $|\mathcal{M}_A|^2$ diverges as $k_t^2\to 0$; in the trace we can therefore approximate $k\approx (1-x)P$ and $p_2\approx \overline{P}$: all the terms excluded by these approximations, in fact, softens the divergence and can be neglected. As a consequence we have:
\begin{equation}\label{MAsquaredapprox}
|\mathcal{M}_A|^2\approx \frac{g^2}{2c_W^2}\,\frac{1}{k_t^2}\,x(1-x)^2Tr\left(\Upsilon\slashed{\overline{P}}
\Upsilon^{*}\slashed{P}\right),
\end{equation}
and the remaining trace is exactly the one obtained in eq. (\ref{bornlevel}) related to the process with no gauge boson emission, showing our first step towards the factorization of the amplitude squared.\\
Considering the interference term $\mathcal{M}_A\mathcal{M}_B^*+
\mathcal{M}_A^*\mathcal{M}_B$ it's possible to use the same trick writing:
\begin{equation}\label{MAMBApprox}
  \mathcal{M}_A\mathcal{M}_B^*\approx \frac{g^2}{2c_W^2}\frac{x^2}{k_t^2}\,Tr\left(\Upsilon\slashed{\overline{P}}
\Upsilon^{*}\slashed{P}\right),
\end{equation}
and as a result we obtain the factorization of the amplitude squared for the three-body decay in the CA respect to the two-body one:
\begin{equation}\label{amplitudefactorized}
|\mathcal{M}_3|^2\approx \frac{g^2}{2c_W^2}\,\frac{x(1+x^2)}{k_t^2}\,|\mathcal{M}_{\rm Born}|^2.
\end{equation}
Referring to the eikonal approximation in eq. (\ref{3gfeynmansudakov2}), we see that using CA it's possible to factorize the amplitude only respect to its squared.

We can apply the CA also considering the three-body phase space of $\Gamma_3$ in eq. (\ref{3body}).
Following \cite{CF} we therefore obtain a complete factorization: the three-body decay width can be expressed as the product of the two-body one times a collinear factor:
\begin{equation}\label{fattorizzazione}
d\Gamma_3\left(Z^{\prime}\to \nu_L\overline{\nu_L}Z\right)\approx d\Gamma_2\left(Z^{\prime}\to \nu_L\overline{\nu_L}\right)\,\frac{\alpha_2}{2\pi}\,\frac{1}{4(1-s_W^2)}\,\frac{1+x^2}{1-x}\,\frac{dxdk_t^2}{k_t^2}.
\end{equation}
Integrating over the final phase space we finally find:
\begin{equation}\label{integroGamma}
d\Gamma_3(x)=\Gamma_2\,\frac{\alpha_2}{2\pi}\,\frac{1}{4(1-s_W^2)}
\,\frac{1+x^2}{1-x}\,\left.L(1-x)\right|_{\rm Sud}\,dx
\end{equation}
with $s=M^2$.

As commonly done in parton models, we interpret the factor multiplying the two-body decay width as the parton distribution for finding a ``neutrino parton'' in the neutrino:
\begin{equation}\label{spectrumpos}
D_{\nu_L\to \nu_L}(x)= \frac{1}{\Gamma_2}\,\frac{d\Gamma_3}{dx}=\frac{\alpha_2}{2\pi}\,\frac{1}{4(1-s_W^2)}\,\frac{1+x^2}{1-x}
\,\left.L(1-x)\right|_{\rm Sud},
\end{equation}
where up to this point this expression take into account just the effects related to the real gauge boson emission. Adding a tree level delta term to account the process without electroweak emission and introducing virtual corrections at the same perturbative order of the real ones we obtain:
\begin{equation}\label{realplusvirtual}
D_{\nu_L\to \nu_L}(x) =\frac{\alpha_2}{2\pi}\,\frac{1}{4(1-2s_W^2)}\,P_{F\to F}
+\delta(1-x)\left\{1+\frac{\alpha_2}{2\pi}\left[\frac{3-2s_W^2}{4(1-s_W^2)}\right]P_{F\to F}^{\rm vir}\right\},
\end{equation}
where:
\begin{equation}\label{Pvirtuale}
P_{F\to F}=\frac{1+x^2}{1-x}\left.L(1-x)\right|_{\rm Sud},
\hspace{1cm}P_{F\to F}^{\rm vir}= \frac{3}{2}\ln\frac{s}{M_W^2}-\frac{1}{2}\ln^2\frac{s}{M_W^2}.
\end{equation}
A complete set of integrated splitting functions is collected in Table \ref{tab:splitting}.
eq.  (\ref{realplusvirtual}) represents a concrete one-loop example,
obtained through a direct
calculation, of the logarithmical structure of the electroweak
fragmentation function $D_{I\to J}$. At this point one might be skeptical about the effective validity of
the CA. In order to remove all
 doubt, we compare in \ref{fullresult} our improved
 CA with the full result of a
 complete three-body calculation done in the context of the Minimal
 Dark Matter model
\cite{MDM}, finding an excellent agreement.
The one-loop fragmentation functions for the entire SM are discussed
into Appendix \ref{frag},
 in the more general context of the electroweak evolution equations
 \cite{EvEq}.

 At this point we are ready to evaluate in this example the energy spectrum
 of stable SM particles produced by the $Z^{\prime}$ decay and for definiteness we consider the case of the neutrino.
We get:
\begin{equation}\label{spettronu}
\frac{dN_{{\rm DM}\to \nu_L}}{dx}=\frac{1}{\Gamma_{\rm tot}}\left\{
\frac{d\Gamma_3(Z^{\prime}\to \nu_L\overline{\nu_L}Z)}{dx}+
\frac{d\Gamma_3(Z^{\prime}\to \nu_Le_L^+W^-)}{dx}
\right\},
\end{equation}
where $\Gamma_{\rm tot}=\Gamma_{ee}+\Gamma_{\nu\nu}=2\Gamma_B$. From eq. (\ref{spectrumpos}) we have:
\begin{equation}\label{peso}
\frac{d\Gamma_3(Z^{\prime}\to \nu_L\overline{\nu_L}Z)}{dx}=
\Gamma_2(Z^{\prime}\to \nu_L\overline{\nu_L})\,D_{\nu_L\to \nu_L}(x)=\Gamma_B\,D_{\nu_L\to \nu_L}(x),
\end{equation}
and, in a similar way:
\begin{equation}\label{peso2}
\frac{d\Gamma_3(Z^{\prime}\to \nu_Le_L^+W^-)}{dx}=
\Gamma_2(Z^{\prime}\to e^+_Le^-_L)\,D_{e^-_L\to \nu_L}(x)=\Gamma_B\,D_{e^-_L\to \nu_L}(x),
\end{equation}
 with $D_{e^-_L\to \nu_L}$ as in (\ref{sys:fermion}); finally we obtain the neutrino spectrum at perturbative order $\alpha_2$:
 \begin{equation}\label{spettronuesempio}
 \frac{dN_{{\rm DM}\to \nu_L}}{dx}=\frac{1}{2}\left\{D_{\nu_L\to \nu_L}(x)+D_{e^-_L\to \nu_L}(x)\right\}.
 \end{equation}

\subsection{Full computation in the Minimal Dark Matter model}\label{fullresult}
In order to validate the eikonal and collinear approximations, we compare its results with a
full computation performed in a specific predictive model.
We consider the weak corrections to DM annihilations  into $W^+W^-$ as predicted by  ``Minimal Dark Matter''  models,
where DM only has electroweak interactions~\cite{MDM}.
This generic situation is realized in the region of the MSSM parameter space
where DM could be the neutral component of the  fermionic  wino weak triplet with a value of the mass $M\sim 3\,\TeV$ dictated by the cosmological DM relic density~\cite{MDM,Hisano:2006nn}.
The same situation can be realized with scalar DM and/or with DM lying in different representations of the weak group.
Particularly interesting are two cases --- a fermionic 5plet and a scalar 7plet with zero hypercharge --- where DM is automatically stable
because SM particles do not have the quantum numbers that could couple to such DM multiplets.
Our computation applies in all such cases: at leading order we have the process DM DM $\to W^+W^-$, and at NLO  the three-body annihilation channels DM DM $\to W^+W^-Z$ and DM DM $\to W^+W^-\gamma$  open up.
 The full expressions for the spectra for the emitted $\gamma$ and $Z$ are quite lengthy, and therefore,
 we write them approximating $M_W \approx M_Z$ and defining $\epsilon\equiv M_W/M \ll 1$,
neglecting terms of  ${\cal O}(\epsilon^2)$:\footnote{Incidentally we find that
at this order the result is the same for both scalar and fermionic DM.
Notice that this Taylor expansion in $\epsilon$ is only valid at $x\gg \epsilon$, and consequently, like the Sudakov approximation,
does not correctly describe the kinematical boundaries.}
\begin{eqnsystem}{sys:4}
\frac{dN_{{\rm DM}\to \gamma}}{dx}&=& \frac{\alpha_{\rm em}}{\pi}\bigg[
   \frac{4 \left(1-x+x^2\right)^2}{(1-x) x}   \ln \frac{2}{\epsilon}
   +\frac{2 \left(4-12x+19x^2-22x^3+20x^4-10x^5+2x^6\right)}{(x-2)^2
   (x-1)x}+ \nonumber \\
&&+   \frac{-6 x^5+32 x^4-74 x^3+84 x^2-48 x+16}{(x-2)^3 (x-1) x}  \ln (1-x)\bigg],\label{eq:gamma} \\
\frac{dN_{{\rm DM}\to Z}}{dx} &=&\frac{\alpha_2}{\pi}\,c_W^2\,\bigg[
\frac{9 x^4-18 x^3+25 x^2-16 x+8}{2 x(1-x)} \ln \left(\frac{2 x/\epsilon}{\sqrt{1-x+x^2} }\right)+  \nonumber \\
&&\frac{2 \left(-3 x^5+16 x^4-37
   x^3+42 x^2-24 x+8\right)}{(2-x)^3 (1-x) x}  \ln (1-x)+ \nonumber \\
   &&-\frac{\left(52-176x+271x^2-247x^3+150x^4-55x^5+9x^6\right) x}{2 (2-x)^2 (1-x) (1-x+x^2)}\bigg],\label{eq:zeta}\\
\frac{dN_{{\rm DM}\to W}}{dx} &=&\delta(1-x)+ \frac{\alpha_2}{\pi}\,c_W^2\,\bigg[
\frac{9 x^4-18 x^3+25 x^2-16 x+8}{4 x(1-x)} \ln \left(\frac{2 x/\epsilon}{\sqrt{1-x+x^2} }\right)+  \nonumber \\
&&(-5-4/x^2+5/x) \ln (1-x)+ \nonumber\\
   &&-\frac{80-224x+425x^2-473x^3+341x^4-140x^5+36x^6}{16 x (1-x) (1-x+x^2)}\bigg]+\\
&&   \frac{\alpha_{\rm em}}{\pi}\,\bigg[
\frac{2 \left(1-x+x^2\right)^2}{x(1-x)} \ln \left(\frac{2 x}{\epsilon}\right)-\frac{30-54x+71x^2-36x^3+12x^4}{6(1-x)x}+  \nonumber \\
&&-\frac{4-7x+6x^2+x^3-4x^4+2x^5}{(1-x)x^2} \ln (1-x)\bigg] \nonumber 
\end{eqnsystem}
In the $W$ spectra the first term is the leading order annihilation, the second one arises from 3-body processes
with an additional $Z$, and the third term from processes with an additional $\gamma$.
Eq.\eq{gamma} agrees with the result computed for fermionic DM in~\cite{Bergstrom:2005ss}.\footnote{Note that the diagram there called ``QED Internal Bremsstrahlung'' in our language is ordinary EW bremsstrahlung from the initial DM and
it does not give any log-enhanced effect because DM is non-relativistic.
We also agree with~\cite{Bergstrom:2005ss} with the terms suppressed by $\Delta M/M_W$, that we do not show because a $\Delta M$ would need a DM coupling to the Higgs,
and consequently extra 2 body annihilations.}

\medskip

\begin{figure}
     \begin{minipage}{0.4\textwidth}
      \centering
       \includegraphics[width=8 cm]{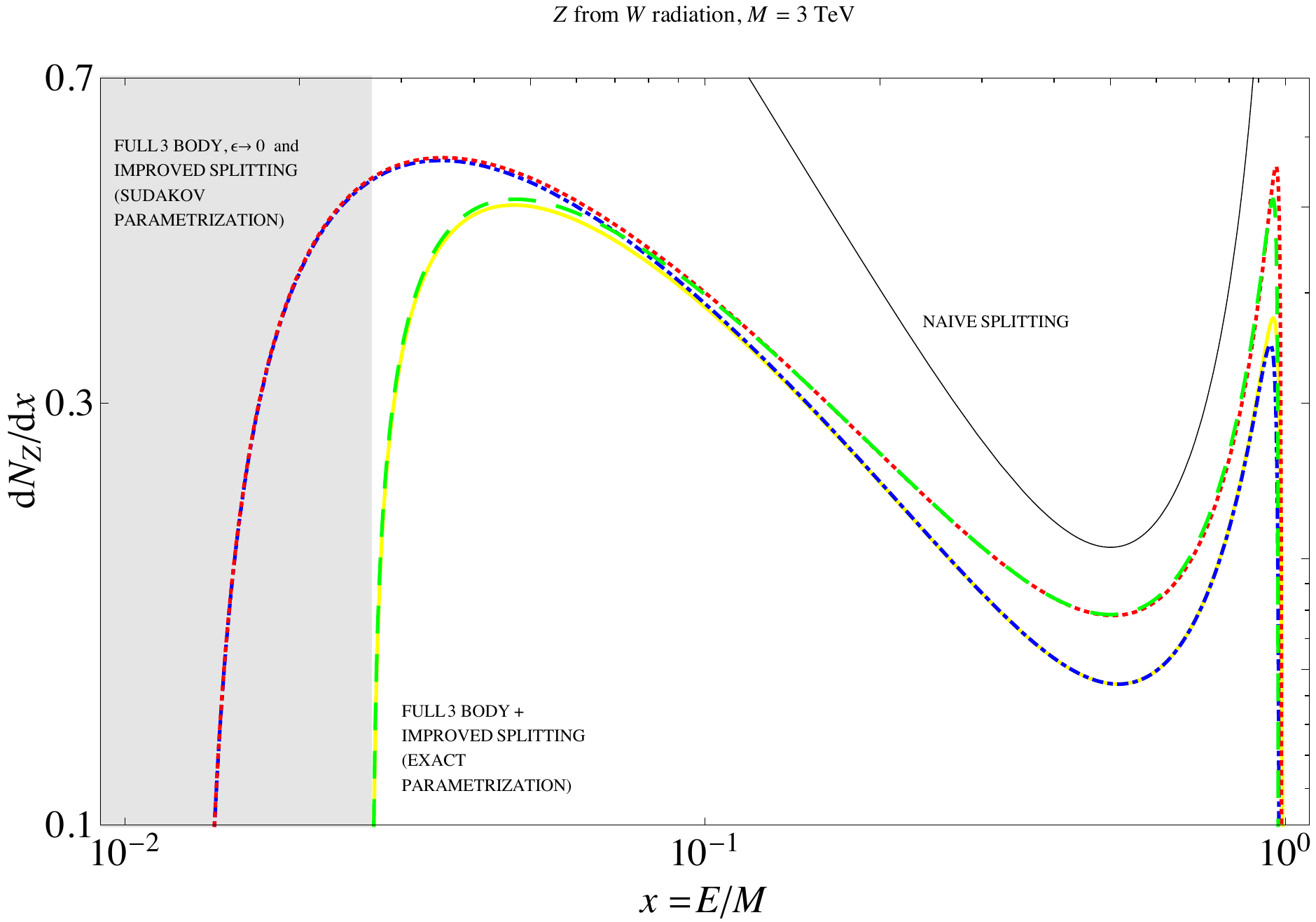}
     \end{minipage}\hspace{1.5 cm}
     \begin{minipage}{0.4\textwidth}
      \centering
       \includegraphics[width=8 cm]{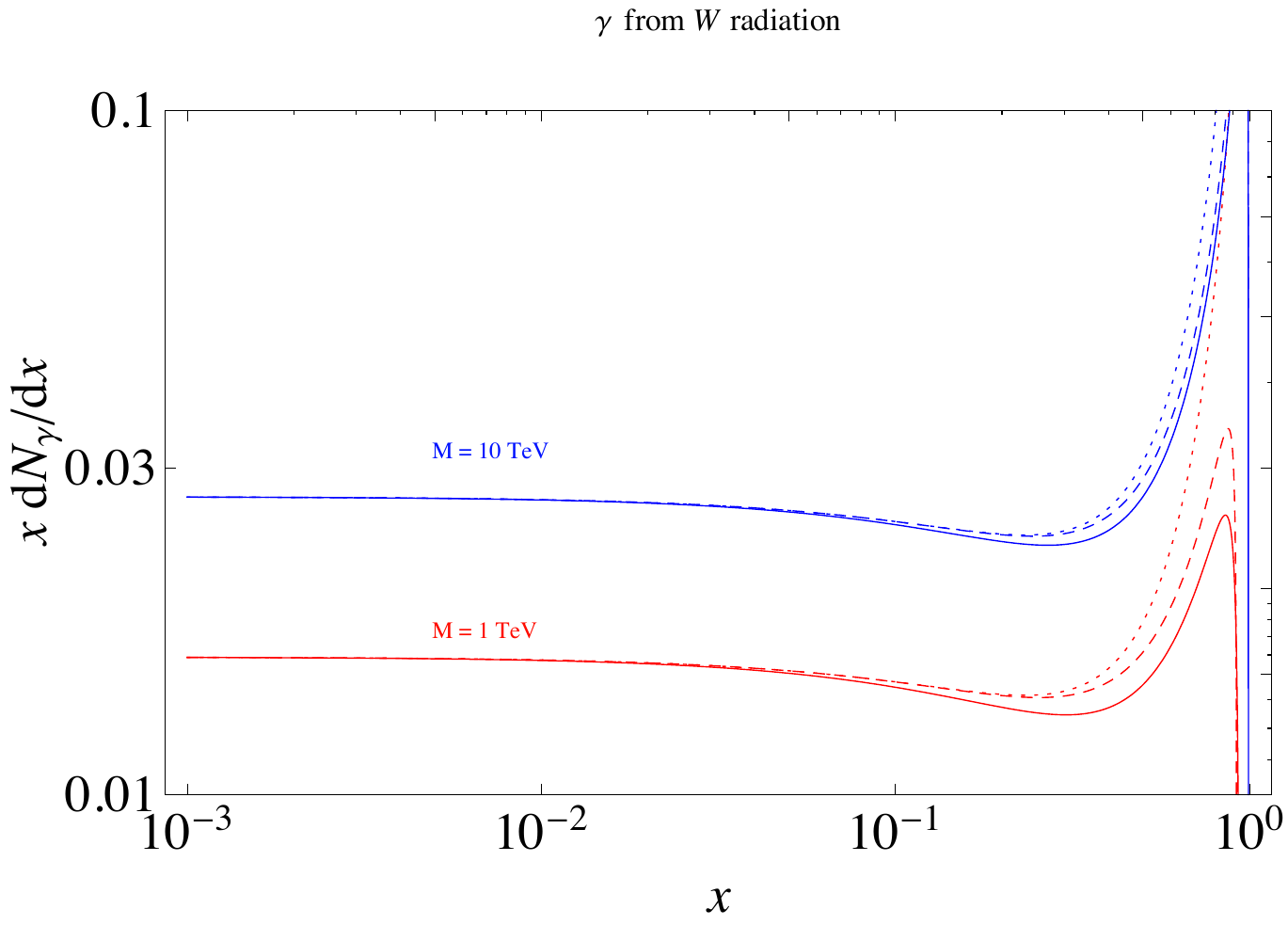}
     \end{minipage}
\caption{\emph{Left, $Z$ from $W$ radiation: Comparison between our full result in the Minimal Dark Matter model (continuous yellow line), with its limit for $\epsilon\equiv M_W/M \to 0$ (blue dot dashed) and with our improved eikonal approximation (red dotted for the Sudakov parametrization and green dashed for the exact one). We show also the comparison with the na\"{\i}ve standard partonic approximation (black continuous line).
Right, $\gamma$ from $W$ radiation: comparison between our full result (continuous red/blue line) with our improved splitting approximation in the exact parametrization (red/blue dashed) and the standard partonic one (red/blue dotted).}}\label{figconfronti}
   \end{figure}

We now compare the full result with its collinear approximation, where the same quantities in Eqs. (\ref{eq:gamma},\ref{eq:zeta}) are described through electroweak fragmentation functions.
In the previous sections we derived the results treating the phase space either within the Sudakov parametrization [Eqs. (\ref{sudakovfinale1},\ref{sudakovfinale2})] or exactly [Eqs. (\ref{exactfinale1},\ref{exactfinale2})]. 
We therefore compare the full $\gamma$ spectrum in eq.~(\ref{eq:gamma}) with:
\begin{equation}\label{gammaemissionSudakov}
\frac{dN_{{\rm DM}\to \gamma}}{dx}=\frac{\alpha_{\rm em}}{\pi}\,2\left[\frac{x}{1-x}L(1-x)
+\frac{1-x}{x}\ln\frac{s}{M_W^2}
+x(1-x)
\ln\frac{s}{M_W^2}\right],
\end{equation}
and the full $Z$ spectrum in eq.~(\ref{eq:zeta}) with:
\begin{equation}\label{ZemissionSudakov}
\frac{dN_{{\rm DM}\to Z}}{dx}=\frac{\alpha_2}{\pi}\,2c_W^2\left[\frac{x}{1-x}L(1-x)
+\frac{1-x}{x}L(x)
+x(1-x)
\ln\frac{s}{M_W^2}\right],
\end{equation}
where $L(x)$ is given in eq.~(\ref{Lxsudakov}) for the Sudakov parametrization and in eq.~(\ref{Lexact}) for the exact phase space. Results are shown in Fig. \ref{figconfronti}, where we depict also the curve corresponding to the na\"{\i}ve partonic approximation, where the upper bound on $\mu^2$ is chosen to be the typical hard scale of the problem $s$, obtaining a universal logarithmical factor $\ln s/M_W^2$ that multiplies the splitting function.

At small values of $x$ the Sudakov parametrization (red dotted line in the left panel) shows a bad behavior compared with the full calculation and don't respect the correct kinematical end-points:
\begin{equation}\label{correctkinematicalendpoints}
\frac{2M_W}{\sqrt{s}}\leq x\leq 1-\frac{2M_W}{\sqrt{s}}.
\end{equation}
This is because in the Sudakov parametrization, considering the splitting process $i\to f+f'$, the variable $x$ don't correspond exactly with the fraction of energy carried by the final particle $f$ respect to its initial value.
On the contrary the exact phase space (green dashed line) gives of course the correct kinematical boundary of the splitting process (\ref{kinendpointsexact}) and shows a correct agreement with the full calculation.

\section{One loop Electroweak Fragmentation Functions}\label{frag}
In the following we collect one loop EW fragmentation functions obtained solving the EW evolution equations in \cite{EvEq} for the entire Standard Model particle spectrum.

EW evolution equations have been already constructed in \cite{EvEq}, exploiting the $\rm{SU}(2)_L$ symmetry, and classifying the states looking to their total isospin quantum numbers; consequently one have to apply a projection technique (explained in details in \cite{EvEq}) in order to convert them to their QCD-like formulation, i.e.\  labeling splittings with particle names:
$D_{i\to j}$ is the single leg fragmentation functions  that encode the probability for a single initial particle $i$ to become a final particle $j$.

Since DM gives particle-antiparticle pairs, to reduce the combinatorics, we combine them into pairs of  primary back-to-back SM particles $I$ instead of a single particle $i$. A simple formula allows us to switch from the `single leg' to the `double leg' convention:
\begin{equation}\label{switch}
D_{I\to J}=c\left(
D_{i\to j}+D_{i\to \overline{j}}+D_{\overline{i}\to j}+D_{\overline{i}\to \overline{j}}
\right),
\end{equation}
where $c=1/2$ for complex final particles (such as the $W$ or the $\nu)$ while $c=1/4$ for real ones (such as the $Z$ or the $\gamma$).



\subsection{Splitting of fermions}
We start considering DM that produces two back-to-back fermions, $f\bar{f}$ with center-of-mass energy $\sqrt{s}$,
and compute the resulting partonic spectrum of other SM particles $A$, $D_{f\to A}(z)$, with $z\equiv E_A/2\sqrt{s}$.
As the $f\bar{f}$ pair produces $A$ and $\bar A$ with the same energy spectrum, we always average over particle and its anti-particle
both in the initial and in the final state, even for real particles such as the $Z$.
The initial fermion can be
$f=\{e_R,e_L,\nu_L,u_L,d_L,u_R,d_R\}$ and is identified by its $T_3=\{-1/2,0,1/2\}$, its electric charge $Q$,
its generation number.
We define the usual coupling to the $Z$, $g_f = T_3 - s_W^2 Q$
and to the photon, $\alpha_{\rm em}=s_W^2 \alpha_2$.
We define the top Yukawa coupling $y_t = m_t/v$ with $v\approx 174\GeV$ and $\alpha_t = y_t^2/4\pi$.
Neglecting all other fermion masses we get:
\begin{eqnsystem}{sys:fermion}
D_{f\to f} (x)&=&\delta(1-x)\bigg[1+\frac{\alpha_2}{2\pi} \;(2\,T_3^2 + \frac{g_f^2}{c_W^2}+ Q^2 \,s_W^2) P_{F\to F}^{\rm vir}\bigg]+\\
&&+ \frac{\alpha_2}{2\pi}\;(\frac{g_f^2}{c_W^2}+ Q^2 s_W^2)\; P_{F\to F}(x), \\
D_{f\to Z_T} (x)&=& \frac{\alpha_2 \;g_f^2}{2\pi c_W^2}\;P_{F\to V}(x),\\
D_{f\to \gamma} (x)&=& \frac{\alpha_{\rm em}}{2\pi}\; Q^2\;P_{F\to V}(x),\\
D_{f\to f'} (x)&=& \frac{\alpha_2}{2\pi} \;2\;T_3^2\;P_{F\to F}(x),\\
D_{f\to W_T}(x) &=& \frac{\alpha_2}{2\pi} \;2\;T_3^2\;P_{F\to V}(x),
\end{eqnsystem}
where $f'$ is the ${\rm SU}(2)_{L}$ partner of $f$ (e.g.\ $f'=e_L$ for $f=\nu_L$ and viceversa).
The virtual term in the first line means that a fraction of the initial $f$ with $x=1$ disappears into $f$ or other particles with $x<1 $.

All splittings of quarks are negligible with respect to the QCD splittings (not written).
For the $t$ and $b$ quarks there are extra splittings (to be summed to the ones listed above due to gauge interactions)
due to the top quark Yukawa interaction, which also have minor effects. Again these splittings depend on the quark polarization;
for simplicity, we write them for average unpolarized $t$, $b$ quarks:
\begin{eqnsystem}{sys:tb}
D_{t\to t}^{\rm Yuk} (x)&=& \delta(1-x)\;\frac{3\,\alpha_t}{8\pi}\; P_{F\to F}^{\rm Yuk,vir}+\frac{\alpha_t}{4\pi} \;P_{F\to F}^{\rm Yuk}(x)\\
D_{t\to b}^{\rm Yuk}(x) &=&D_{b\to t}^{\rm Yuk}(x)=\frac{\alpha_t}{8\pi} \;P_{F\to F}^{\rm Yuk}(x)\\
D_{t\to Z_L}^{\rm Yuk} (x)&=&D_{t\to h}^{\rm Yuk} =D_{t\to W_L}^{\rm Yuk}=D_{b\to W_L}^{\rm Yuk}=\frac{\alpha_t}{8\pi}\; P_{F\to S}^{\rm Yuk}(x)\\
D_{b\to b}^{\rm Yuk}(x) &=& \delta(1-x)\;\frac{\alpha_t}{8\pi}\; P_{F\to F}^{\rm Yuk,vir}.
\end{eqnsystem}


\subsection{Splitting of Higgses}
The Higgs doublet $H$ contains the physical Higgs $h$, as well as the goldstone components that describe the longitudinal
polarizations $W_L$ and $Z_L$ of the SM massive vectors. Splittings induced by the top quark Yukawa coupling have a significant effect;
we describe them without specifying the polarizations of the $t$, $b$ quarks, which make a negligible difference.

For the physical Higgs $h$ we have:
\begin{eqnsystem}{sys:h}
D_{h\to h}(x) &=& \delta(1-x)\bigg[1+\frac{3\;\alpha_2+\alpha_Y}{2\pi\cdot 4}  P_{S\to S}^{\rm vir}\bigg]
+\frac{3\;\alpha_t}{4\pi}P_{S\to F}^{\rm vir,Yuk}, \\
D_{h\to W_T}(x) &=& \frac{\alpha_2 }{2\pi}\; \frac{1}{2} \; P_{S\to V}(x) = 2\;c_W^2 \;D_{h\to Z_T}(x),\\
D_{h\to W_L}(x) &=& \frac{\alpha_2 }{2\pi}\; \frac{1}{2}\;  P_{S\to S}(x) = 2\; c_W^2 \; D_{h\to Z_L}(x),\\
D_{h\to t}(x) &=&\frac{3\; \alpha_t}{2\pi} \; P_{S\to F}^{\rm Yuk}(x).
\end{eqnsystem}
The same expressions hold for an initial $Z_L$. For the longitudinal $W_L$ we have:
\begin{eqnsystem}{sys:WL}
D_{W_L\to W_L}(x) &=& \delta(1-x)\bigg[1+\frac{3\; \alpha_2+\alpha_Y}{2\pi\cdot 4}  P_{S\to S}^{\rm vir}\bigg] +
\frac{\alpha_2+\alpha_Y}{2\pi\cdot 4} P_{S\to S}(x)+\frac{3\; \alpha_t}{4\pi}P_{S\to F}^{\rm vir,Yuk},\\
&=&  \delta(1-x)\bigg[1+\frac{\alpha_{\rm em}}{2\pi}\;  P_{S\to S}^{\rm vir}\bigg] +
\frac{\alpha_{\rm em}}{2\pi}\; P_{S\to S}(x) +\cdots,\label{eq:QED}\\
D_{W_L\to h}(x) &=& \frac{\alpha_2 }{2\pi}\;\frac{1}{4}\; P_{S\to S}(x) = D_{W_L\to Z_L}(x),\\
D_{W_L\to Z_T}(x) &=& \frac{\alpha_2 }{2\pi}\;\frac{g_{e_L}^2}{c_W^2} \;P_{S\to V}(x), \\
D_{W_L\to \gamma}(x) &=& \frac{\alpha_{\rm em}}{2\pi}\;P_{S\to V}(x),\\
D_{W_L\to t}(x) &=&D_{W_L \to b}(x)=\frac{3\;\alpha_t}{4\pi}\; P_{S\to F}^{\rm Yuk}(x).
\end{eqnsystem}
Eq. \eq{QED} shows the QED corrections only, in case one needs to subtract them.

\subsection{Splitting of vectors}
For the transverse $W^\pm$ we find:
\begin{eqnsystem}{sys:WT}
D_{W_T\to W_T}(x) &=& \delta(1-x)\bigg[1+\frac{\alpha_2}{2\pi} \;2\; P_{\rm SU(2)}^{\rm vir} \bigg]+\frac{\alpha_2}{2\pi}\;P_{V\to V}(x), \\
D_{W_T\to Z_T}(x) &=& \frac{\alpha_2 }{2\pi}\;c_W^2\;P_{V\to V}(x),\\
D_{W_T\to \gamma}(x) &=& \frac{\alpha_2 }{2\pi}\;s_W^2\;P_{V\to V}(x),\\
D_{W_T\to f_L}(x) &=&\frac{\alpha_2 }{2\pi}\;\frac{1}{2}\;N_c\; P_{V\to F}(x),\qquad f=\{e,\nu_e,d,u; \mu,\nu_\mu, s,c; \tau,\nu_\tau,b,t\},\\
D_{W_T\to h}(x) &=& \frac{\alpha_2}{2\pi}  \;\frac{1}{4} \;P_{V\to S}(x)=D_{W_T\to Z_L}(x),\\
D_{W_T\to W_L}(x) &=& \frac{\alpha_2}{2\pi}\;  \frac{1}{2}\; P_{V\to S}(x),
\end{eqnsystem}
and for the transverse $Z$ we find:
\begin{eqnsystem}{sys:ZT}
D_{Z_T\to Z_T}(x) &=& \delta(1-x)\bigg[1+\frac{\alpha_2}{2\pi}\; \bigg(2\;c_W^2 \;P_{\rm SU(2)}^{\rm vir} + \frac{s_W^4}{c_W^2}\; P_{\rm U(1)}^{\rm vir}\bigg)\bigg],\\
D_{Z_T\to W_T}(x) &=& \frac{\alpha_2 }{2\pi}\;2\;c_W^2\;P_{V\to V}(x),\\
D_{Z_T\to f}(x) &=&\frac{\alpha_2 g_f^2}{2\pi c_W^2}\;2\;N_c \;P_{V\to F}(x),\qquad f=\{\nu_e,e_L,e_R,u_L,u_R,d_L,d_R,\ldots\},
\\
D_{Z_T\to h}(x) &=& \frac{\alpha_2\; g_\nu^2}{2\pi c_W^2}\;P_{V\to S}(x) =D_{Z_T\to Z_L}(x),\\
D_{Z_T\to W_L}(x) &=& \frac{\alpha_2}{2\pi} \; \frac{2 \;g_{e_L}^2}{c_W^2} \;P_{V\to S}(x),
\end{eqnsystem}
where we defined $N_c=1~(3)$ if $f$ is a lepton (a quark) doublet; $N_{\rm gen}=3$, and:
\begin{eqnsystem}{sys:Pvit}
P_{\rm SU(2)}^{\rm vir} &=&\frac{1}{2}P_{V\to V}^{\rm vir}+\frac{1}{4}P_{V\to S}^{\rm vir}+N_{\rm gen} \;P_{V\to F}^{\rm vir},\\
P_{\rm U(1)}^{\rm vir} &=& N_{\rm gen} \;
(2Y_L^2+Y_E^2+6Y_Q^2+3Y_U^2+3Y_D^2) \;P_{V\to F}^{\rm vir}+2\;Y_L^2 \;P_{V\to S}^{\rm vir}.
\end{eqnsystem}
The $\gamma$ contributes to $P_{W_T\to W_T}$ and can be excluded by dropping $1 = c_W^2 + s_W^2\to c_W^2$ in front of
$P_{V\to V}$, both real and virtual. (Notice that PYTHIA does not include QED radiation from $W^\pm$).

Finally for the photon we have:
\begin{eqnsystem}{sys:gamma}
D_{\gamma \to \gamma}(x) &=& \delta(1-x)\bigg[1+\frac{\alpha_{\rm em}}{2\pi} \bigg(2\; P_{\rm SU(2)}^{\rm vir} +  P_{\rm U(1)}^{\rm vir}\bigg)\bigg],\\
D_{\gamma \to W_T}(x) &=& \frac{\alpha_{\rm em}}{2\pi} \;2\; P_{V\to V}(x),\\
D_{\gamma \to W_L}(x) &=& \frac{\alpha_{\rm em}}{2\pi} \;2 \;P_{V\to S}(x),\\
D_{\gamma \to f}(x) &=&\frac{\alpha_{\rm em}}{2\pi}\;2Q^2\;N_c \;P_{V\to F}(x),\qquad f=\{e_L,e_R,u_L,u_R,d_L,d_R,\ldots\}.
\end{eqnsystem}
%
%
%
%


\def\np#1#2#3{{\sl Nucl.~Phys.\/}~{\bf B#1} {(#2) #3}} \def\spj#1#2#3{{\sl
Sov.~Phys.~JETP\/}~{\bf #1} {(#2) #3}} \def\plb#1#2#3{{\sl Phys.~Lett.\/}~{\bf
B#1} {(#2) #3}} \def\pl#1#2#3{{\sl Phys.~Lett.\/}~{\bf #1} {(#2) #3}}
\def\prd#1#2#3{{\sl Phys.~Rev.\/}~{\bf D#1} {(#2) #3}} \def\pr#1#2#3{{\sl
Phys.~Rep.\/}~{\bf #1} {(#2) #3}} \def\epjc#1#2#3{{\sl Eur.~Phys.~J.\/}~{\bf
C#1} {(#2) #3}} \def\ijmp#1#2#3{{\sl Int.~J.~Mod.~Phys.\/}~{\bf A#1} {(#2) #3}}
\def\ptps#1#2#3{{\sl Prog.~Theor.~Phys.~Suppl.\/}~{\bf #1} {(#2) #3}}
\def\npps#1#2#3{{\sl Nucl.~Phys.~Proc.~Suppl.\/}~{\bf #1} {(#2) #3}}
\def\sjnp#1#2#3{{\sl Sov.~J.~Nucl.~Phys.\/}~{\bf #1} {(#2) #3}}
\def\hepph#1{{\sl hep--ph}/{#1}}

\end{document}